\documentclass[%
 aip,
 pop,
 amsmath,amssymb,
 reprint,%
]{revtex4-1}

\usepackage{graphicx}
\usepackage{dcolumn}
\usepackage{bm}
\usepackage{color}

\usepackage[utf8]{inputenc}
\usepackage[T1]{fontenc}
\usepackage{mathptmx}
\usepackage{etoolbox}

\makeatletter
\def\@email#1#2{%
 \endgroup
 \patchcmd{\titleblock@produce}
  {\frontmatter@RRAPformat}
  {\frontmatter@RRAPformat{\produce@RRAP{*#1\href{mailto:#2}{#2}}}\frontmatter@RRAPformat}
  {}{}
}%
\makeatother
\begin{document}

\preprint{AIP/123-QED}

\title[Butterfly Effect in Plasma]{Butterfly Effect in Kinetic Plasma Turbulence}
\author{L. Henderson}

\email{ljhenderson3@wisc.edu}
\author{V. Zhdankin}%

\affiliation{ 
Department of Physics, University of Wisconsin-Madison, Madison, WI 53706 
}%

\date{\today}

\begin{abstract}
The butterfly effect is a fundamental process in chaotic systems, but it has not previously been characterized in detail for a collisionless turbulent plasma. In this work, we apply particle-in-cell (PIC) simulations of driven turbulence to study the emergence of the butterfly effect from a small initial perturbation to the simulated particles. In particular, we compare pairs of (baseline and perturbed) simulations that differ only by a localized initial perturbation (down to the single-particle level). We find that the effect of the localized perturbation spreads radially throughout the PIC simulation, causing a difference between each simulation pair that is exponentially amplified over time within the affected region. The fields, including the electric current density, eventually become statistically decorrelated. We construct an analytical model combining spatial expansion with exponential amplification to approximate the early-time evolution. Varying the initial perturbation primarily affects the early-time evolution, while having little effect on the final, decorrelated state. We describe parameter scans with varying plasma beta and temperature (covering sub-relativistic and ultra-relativistic regimes). Our results show that remarkably small perturbations will grow to decorrelate each pair of simulations within a few dynamical times; the imposed perturbation can no longer be resolved once it falls below the numerical precision floor set by the machine precision. These results confirm that the butterfly effect is very strong and robust in a collisionless plasma, with implications for the predictability of turbulent plasma systems. 
\end{abstract}

\maketitle

\section{\label{sec:level1}Introduction}

In a 1972 presentation, Edward Lorenz posed the question "does the flap of a butterfly's wings in Brazil set off a tornado in Texas?" \cite{lorenz1972predictability}. The question, more a philosophical demonstration than an in-depth study, explores the timescale of stability in the atmosphere among other turbulent systems, illustrating the idea of how a small, seemingly insignificant change could lead to an unpredictable outcome. This concept, commonly known as the "butterfly effect," arises naturally in turbulent systems where small changes in the initial conditions of a system can grow exponentially over time, making long-term predictions impossible \cite{Palmer_2014}. Earlier on in 1962, Lorenz numerically solved a simple system representing cellular convection, finding all solutions to be unstable, showing that slightly different initial conditions can produce drastically different results \cite{lorenz1963deterministic}. Expanding on this, in 1968 Lorenz demonstrated that in certain fluid systems, two states differing initially by only a small observational error would evolve into completely uncorrelated states within a finite time interval, finding an intrinsic finite range of predictability that can be applied to Earth's atmosphere \cite{lorenz1969predictability}. The extent of the butterfly effect has been explored extensively in hydrodynamical turbulence \citep[e.g.,][]{mccauley1990introduction, washington_2000, manneville_2010} and atmospheric systems \cite{lorenz1963deterministic, shen_etal_2022}; however, its implications for plasmas remain mostly unexplored. 

Kinetic plasma turbulence offers a unique setting for studying the nature of the butterfly effect. Here, kinetic theory is used to model collisionless plasmas that can exist in highly nonthermal states. Perturbations of the particle distribution function serve as the analogue of the butterfly in Lorenz's posed question. Naively, one may expect that dissipative processes and magnetic self-organization may act to damp out the perturbation, while turbulent diffusion and instabilities will amplify it. The interplay between these competing processes leads to a rich problem where the butterfly effect may be more nuanced than in conventional systems.

In this work, we use particle-in-cell (PIC) simulations to investigate the butterfly effect in a turbulent plasma from first principles. Numerical simulations remain a very practical way to investigate the sensitivity of plasma dynamics to small perturbations with high precision \cite{Verboncoeur_2005, Birdsall_1991}. In particular, PIC simulations are a standard numerical technique for studying kinetic plasma physics, amenable to the inclusion of small perturbations, as needed for controlled studies of the butterfly effect. Specifically, our research aims to determine whether small perturbations at the simulated particle level can trigger macroscopic unpredictability similar to that observed in other chaotic systems. To investigate this, we consider idealized PIC simulations of driven kinetic turbulence, in both sub-relativistic and ultra-relativistic plasma regimes. 

Throughout this numerical investigation, 
our central questions are:
\begin{enumerate}
    \item How does an initially localized perturbation grow (both qualitatively and quantitatively) in a kinetic plasma turbulence simulation?
    \item Do two simulations that initially differ only by a slight perturbation eventually become statistically decorrelated, and if so, over what characteristic timescale?
    \item How do physical parameters such as plasma beta, temperature (including sub-relativistic and ultra-relativistic regimes), and perturbation size affect the rate of decorrelation between the two simulations?
\end{enumerate}

We find that initially localized perturbations grow to produce macroscopic differences throughout the turbulent plasma, even if the perturbation is at the level of an individual particle. The perturbation spreads spatially while its magnitude increases within the affected region, eventually driving the baseline and perturbed systems toward statistical decorrelation. The differences saturate within a few light-crossing times, regardless of the initial parameters. The fitted early-time growth rate, however, is considerably larger (in units of light-crossing times) in the ultra-relativistic regime than in the sub-relativistic regime. Comparisons among sub-relativistic simulations with different values of plasma beta and temperature demonstrate that the measured growth depends on the specific plasma physical regime, although these scans do not isolate the mechanisms responsible for that dependence. The measured differences include the contributions from the amplification of numerical background fluctuations, limiting the smallest imposed perturbations that can be reliably distinguished.

\section{Analytical theory} 

In this section, we describe basic analytical expectations for how the butterfly effect may unfold in a turbulent plasma. This will set the stage for the numerical observation of the butterfly effect in our PIC simulations (Secs.~\ref{sec:simulations} and \ref{sec:results}).

\subsection{Physical system configuration \label{sec:phys_sys_config}}

In this paper, we consider the manifestation of the butterfly effect in a collisionless plasma that experiences driven turbulence. For simplicity, we focus on a plasma that is two-dimensional in position space but three-dimensional in momentum space. We also focus on electron-positron (pair) plasma, which avoids the complexity of distinguishing between perturbations in electrons and ions. We assume that the system is in an initial uniform equilibrium of temperature $T_0$ (including thermal or numerical noise) and particle number density $n_0$, with a constant magnetic field $\boldsymbol{B}_0 = B_0 \hat{\boldsymbol{z}}$ that is directed perpendicular to the $xy$ domain. Turbulence is then driven from this initial state by an external force.

Before proceeding, we note that there are several characteristic velocities in a collisionless plasma, which may control the speed at which a perturbation spreads spatially. The fastest velocity is the speed of light $c$; however, for a sub-relativistic system, the particles will evolve on a much slower timescale. The Alfv\'{e}n velocity $v_A = B_0/(4 \pi \rho)^{1/2}$ describes transverse waves propagating along the magnetic field. Here, $\rho$ is the mass density and we assume a sub-relativistic plasma. Since $\boldsymbol{B}_0$ is perpendicular to the $xy$ domain, Alfv\'{e}n waves are negligible at early times. Instead, fast magnetosonic modes can more efficiently transfer information through the system. The sub-relativistic fast mode velocity is given by
\begin{equation}
v_f^2 = c_s^2+v_A^2 \, ,
\end{equation}
where $c_s = (\Gamma T_0/m)^{1/2}$ is the speed of sound, $\Gamma$ is the adiabatic index, and $m$ is the particle rest mass. In our case, we anticipate that $\Gamma=2$ due to the double-adiabatic equation of state for collisionless plasma with correlated density and magnetic fluctuations\citep{chew_goldberger_low_1956}, which causes the plasma to act like a collisional 2D ideal gas in the direction perpendicular to the mean magnetic field. Introducing the plasma beta as the ratio of thermal to magnetic pressure $\beta = 8\pi n_0 T_0/B_0^2$, we can then express $c_s = (\Gamma/2)^{1/2} \beta^{1/2} v_A$. For $\beta = 1$ and $\Gamma=2$, we then simply have that $v_f \approx 2^{1/2} c_s$. For the case of an ultra-relativistically hot plasma, $v_A$ must be corrected to use the relativistic enthalpy rather than rest mass. In this limit, fast modes will travel at close to the speed of light \citep{ugarov_etal_2026}. 

We compare two systems which are identical apart from the presence of a small localized perturbation in the initial conditions of the second case. We refer to the first system as the "baseline" case and the second system as the "perturbed" case. We can model the expansion of the perturbation between the two systems as follows. We suppose that any small perturbation is amplified exponentially in time at a constant Lyapunov-like growth rate, $\lambda$ (in dimensions of inverse time) \citep{pikovsky_politi_2016}. In addition, the spatial area that this perturbation grows over increases with time due to the propagation of information at some constant characteristic speed $v_0$ (which could be, for example, the speed of light $c$ or sound speed $c_s$). We will refer to the boundary of the region being influenced by the perturbation as the "information front" (which is usually the region constrained by causality). We additionally refer to the boundary of the region containing the primary, high-amplitude perturbation as the "propagation front", which may lag behind the information front. 

Mathematically, the overall growth of the perturbation can be described as follows. The two systems can be characterized at macroscopic scales by their fields and fluid quantities; in the present case, we consider their electric current densities $\boldsymbol{J}_1(\boldsymbol{r},t)$ and $\boldsymbol{J}_2(\boldsymbol{r},t)$, respectively. We then define the difference in current densities as $\Delta \boldsymbol{J}(\boldsymbol{r},t) = \boldsymbol{J}_1(\boldsymbol{r},t) - \boldsymbol{J}_2(\boldsymbol{r},t)$, and the rms current density difference as
\begin{equation}
    |\Delta \boldsymbol{J}|^2_{\rm rms}(t) = \frac{1}{A_{\rm tot}} \int \, |\Delta \boldsymbol{J}(\boldsymbol{r},t)|^2\; d^2r \, , \label{jrms_def_new}
\end{equation}
where $A_{\rm tot} = \int d^2r$ is the total area of the system. For a square domain of side length $L$ (as will be considered in the PIC simulations),  $A_{\rm tot}=L^2$.

In addition to $\Delta \boldsymbol{J}$, we can define differences in a way analogous to Eq.~\ref{jrms_def_new} for any other physical quantities that characterize the system, such as magnetic field, electric field, density, pressure, etc. In this paper, we limit ourselves to considering $\Delta \boldsymbol{J}$ and the difference in magnetic fields, $\Delta \boldsymbol{B}$, for simplicity. We anticipate that other quantities will exhibit qualitatively similar trends to these, since they are all coupled dynamically.

\subsection{Dimensionless physical parameters}

Both the normalized temperature $T_0/(mc^2)$ and plasma beta $\beta$ are expected to influence the propagation and amplification of a perturbation. In the sub-relativistic regime, increasing the temperature raises the sound speed and may therefore increase the effective propagation velocity $v_0$, allowing the perturbed region to occupy a larger fraction of the domain in a given physical time. This can increase $|\Delta \boldsymbol{J}|^2_{\rm rms}$, although it does not by itself imply a larger growth rate $\lambda$. Plasma beta controls the relative importance of particle and magnetic dynamics. Changes in $\beta$ may therefore alter both the plasma response to the perturbation and its measured growth.

\subsection{Dependence on initial perturbation angle \label{angular_dependence_model}}

So far, we have not specified the particular form of the perturbation. We will introduce initial perturbations by rotating the particle momenta by an angle $\Delta \theta$ for both particle species within a localized circular region. This rotation is applied counter-clockwise around $\hat{\boldsymbol{z}}$ to the vector components $p_x$ and $p_y$ within the 2D plane, with the out-of-plane momentum $p_z$ kept fixed. 
The change in the local current density caused by rotating a single particle of charge $q$ is
\begin{align}
    \Delta\boldsymbol{J} \propto q \Delta \boldsymbol{v} \, ,
\end{align}
where $\Delta \boldsymbol{v}=\boldsymbol{v}_1-\boldsymbol{v}_2$ is the velocity perturbation of the particle between the two systems. If we choose $\Delta \boldsymbol{v}$ to be a pure rotation of angle $\Delta\theta$, so that $|\boldsymbol{v}_1| = |\boldsymbol{v}_2| \equiv v$, then
\begin{align}
    |\Delta\boldsymbol{J}| &\propto |q| \sqrt{|\boldsymbol{v}_1|^2 - 2 \boldsymbol{v}_1 \cdot \boldsymbol{v}_2 + |\boldsymbol{v}_2|^2 } \nonumber \\
    &\propto |q|v \sqrt{2-2\cos{(\Delta \theta)}} \nonumber \\
    &\propto 2 |q v  \sin{(\Delta \theta/2)}| \, .
\end{align}
Therefore, for small angles, a Taylor expansion gives $|\Delta \boldsymbol{J}| \propto \Delta\theta$. This leads us to anticipate that the initial rotational perturbation controls the initial rms current density difference, 
\begin{equation}
    |\Delta \boldsymbol{J}|_{\rm rms} (t=0) \propto \Delta\theta \, . \label{eq:jchartheta}
\end{equation}
Note that in a PIC simulation, the perturbation is applied to numerical "macroparticles" rather than physical particles (of which there would be a much greater number). The coefficient of proportionality in Eq.~\ref{eq:jchartheta} will then depend on the numerical parameters of the PIC simulation, in particular, the number of macroparticles per cell and the cell size.

\subsection{Growth of early-time perturbations \label{early_time_pert}}

The initial growth of the perturbation occurs due to the spread of information by thermal (or numerical) noise. This is the fastest timescale if turbulence is gradually driven from a uniform state (as we assume) and the turbulence is sub-sonic. 

Suppose that at $t=0$, we have a circular perturbed region of radius $r_0$ at the origin, corresponding to an area $A(t=0) = \pi r_0^2$. We assume that all the area outside of the perturbed circular region has $\Delta \boldsymbol{J} = 0$. We suppose that the small perturbation grows exponentially within the circular region of radius $r = r_0 + v_0 t$ due to the non-linear dynamics, such that the perturbed current can be described by
\begin{equation}
    |\Delta \boldsymbol{J}(\boldsymbol{r},t)| = \begin{cases} \delta J_0 e^{\lambda (t-t_0)} \, , & \text{ if $r < r_0 + v_0 t$} \, \\
    0 \, , & \text{ otherwise} 
    \end{cases} \, , \label{exponential_bit}
\end{equation}
where $\delta J_0$ is the amplitude of the initial perturbation and $t_0 = r/v_0$. Eq.~\ref{exponential_bit} represents the exponential growth of the perturbed plasma inside the region of area $A(t)$, accounting for the time offset that it takes for the initial perturbation to grow to that radius. So, as time evolves, we maintain this exponential growth, but the perturbation size itself expands with some constant speed $v_0$.

We combine Eqs.~\ref{jrms_def_new} and \ref{exponential_bit}, along with the assumption that $r_0 \lambda/v_0 \ll 1$, to obtain
\begin{align}
|\Delta \boldsymbol{J}|^2_{\rm rms}(t) &= \frac{1}{A_{\rm tot}} \int_0^{v_0 t} (\delta J_0)^2 e^{2\lambda (t - r/v_0)} 2 \pi r \, dr \nonumber \\
&= \frac{\pi (\delta J_0)^2 v_0^2}{2 A_{\rm tot} \lambda^2} \left( e^{2 \lambda t} - 2 \lambda t - 1 \right) \, . \label{jrms_eval}
\end{align}
Absorbing the constants into the term $J_{\rm char} \equiv \sqrt{\pi }\delta J_0 v_0/(\sqrt{2 A_{\rm tot}} \lambda)$ that will be used as a fit parameter, we can write Eq.~\ref{jrms_eval} as
\begin{align}
    |\Delta \boldsymbol{J}|_{\rm rms}(t) = J_{\rm char} \left(e^{2\lambda t}-2\lambda t-1 \right)^{1/2} \, . \label{eq:fit}
\end{align}
When $\lambda t \gg 1$, Eq.~\ref{eq:fit} approaches the simple exponential form $J_{\rm char}e^{\lambda t}$, while for $\lambda t \ll 1$ it approaches $2^{1/2} J_{\rm char} \lambda t$. 

We will use Eq.~\ref{eq:fit} to fit the PIC simulation data in the sub-relativistic simulations (in Sec.~\ref{sec:results}) and ultra-relativistic simulations (in Sec.~\ref{sec:ur_results}).

\subsection{Intermediate time evolution\label{int_time_evolution}}

At a time of around $t\sim L/(2v_0)$, where $L$ is the system size along one axis, the boundary of the expanding perturbed region first reaches the nearest boundary. Beyond this time, the circular expansion assumed in Sec.~\ref{early_time_pert} no longer describes the geometry of the affected region. Despite this, the perturbation will continue to be amplified before approaching late-time saturation. Thus, the evolution enters a new "intermediate" stage, where $|\Delta \boldsymbol{J}|_{\rm rms}$ continues to increase, but its spatial expansion and amplitude growth can no longer be described using the early-time model. At some point, we anticipate that the growth of the perturbation becomes dominated by the turbulent dynamics rather than the noise.  Turbulence is known to amplify initially small differences and limit predictability in hydrodynamic systems \citep{PhysRevLett.119.054102}, but developing a corresponding quantitative theory for this stage in a kinetic plasma is beyond the scope of the present work. We analyze this intermediate stage and describe its measured evolution empirically in Sec.~\ref{quant_current_diff}.

\subsection{Late time saturation \label{latetimemodel}}

If the butterfly effect is realized, then after sufficiently long times, the perturbation will grow to have a strong influence on every part of the system. At this stage, we would expect the two reference systems to become completely uncorrelated, as if they were initialized from completely different random seeds.

We can write Eq.~\ref{jrms_def_new} as
\begin{align}
    |\Delta \boldsymbol{J}|^2_{\rm rms} &= \frac{1}{A_{\rm tot}} \int |\boldsymbol{J}_1-\boldsymbol{J}_2|^2 \, d^2r \nonumber \\
    &= \frac{1}{A_{\rm tot}} \int \left( |\boldsymbol{J}_1|^2 - 2 \boldsymbol{J}_1 \cdot \boldsymbol{J}_2 + |\boldsymbol{J}_2|^2 \right) \, d^2r \, . \label{eq:jrms_expanded}
\end{align}
When the two simulations are uncorrelated, then the term $\int \boldsymbol{J}_1 \cdot \boldsymbol{J}_2 \, d^2r = 0$ by symmetry. In addition, since the two simulations have identical parameters apart from the perturbation, we expect that $\int |\boldsymbol{J}_1|^2 \, d^2r \approx \int |\boldsymbol{J}_2|^2 \, d^2r \equiv A_{\rm tot} J_{\rm rms}^2$, where $J_{\rm rms}(t)$ is the rms current density for either system individually at the corresponding time. Then Eq.~\ref{eq:jrms_expanded} becomes
\begin{align}
|\Delta \boldsymbol{J}|_{\rm rms}(t) \approx \sqrt{2} J_{\rm rms}(t) \, .
\end{align}
This gives the expected saturation value of the rms current density difference, and defines the point where we refer to the two simulations as "decorrelated." We do not expect $|\Delta \boldsymbol{J}|_{\rm rms}$ to exceed this value, since doing so would require unphysical anti-correlations.

\section{Numerical methods} \label{sec:simulations}

\subsection{PIC simulation setup}

We performed a series of PIC simulations with the code {\em Zeltron} \citep{Cerutti_2013}. The numerical configuration is similar to our previous work\cite{Zhdankin_2021} but reduced to two spatial dimensions: thus, the simulations are two-dimensional in position space but retain all three components of the particle momenta and electromagnetic fields. The domain is a square with length $L$ in each direction, and periodic boundary conditions are imposed in both the $x$ and $y$ directions. The plasma consists of electron-positron pairs randomly sampled from an initial Maxwell-J\"uttner distribution of temperature $T_0$ and total number density $n_0$. The system has a uniform mean magnetic field $\boldsymbol{B}_0 = B_0 \hat{\boldsymbol{z}}$. In addition, we initialize a fluctuating component to the magnetic field to facilitate the development of turbulence, so that the total magnetic field profile is
\begin{align}
\boldsymbol{B}= \frac{1}{2} B_0 \cos{\left( 2 \pi y/L \right)} \hat{\boldsymbol{x}} + \frac{1}{2} B_0 \cos{\left( 2 \pi x/L \right)} \hat{\boldsymbol{y}} + B_0 \hat{\boldsymbol{z}} \, . \label{eq:ic}
\end{align}

An incompressible random external force is used to drive modes at large scales (with wavenumbers $|\boldsymbol{k}| \le 8^{1/2} 2\pi/L$), with strength tuned to make the rms amplitude of magnetic fluctuations comparable to the mean field, $\delta B_{\rm rms} \approx B_0$. Numerical parameters are $N_{\rm ppc} = 32$ particles per cell per species and a resolution of 12 cells per initial gyroradius, $\Delta x = \Delta y = \rho_{e0}/12$ (which allows us to accurately resolve the Debye length for the lowest temperatures). For non-relativistic simulations, we use $\rho_{e0} \approx (3 m_e T_0)^{1/2}/eB_0$ (where $m_e$ is the electron rest mass), while for ultra-relativistic simulations, we use $\rho_{e0} \approx 3 T_0/e B_0$. The box size is chosen to be $L = 128 \rho_{e0} = 1536 \Delta x$. The timestep of the simulations is set by the Courant-Friedrichs-Lewy condition, $\Delta t \approx 2^{-1/2} \Delta x/c$. The PIC simulations include charge and current filtering to reduce numerical noise; we confirmed that this numerical optimization does not influence the results. 

To produce the butterfly effect, we require precise reproducibility of the initial conditions in the PIC simulations. Thus, we use a fixed random-number seed to generate every simulation. This causes particles in all simulations to be initialized with the same set of positions and momenta. Additionally, the exact same external driving (with random phases determined by the seed) is applied to each baseline-perturbed simulation pair. We confirmed that under these conditions, two initially identical simulations exhibit no differences over the duration of the simulation.

Each comparison consists of a baseline simulation and a perturbed simulation initialized with otherwise identical numerical and physical parameters. The only initial difference between the paired simulations is a rotation of the initial particle momenta within a prescribed circular region of radius $r_0$ around the center of the simulation domain; i.e. for particles with $(x,y)$ positions satisfying
\begin{equation}
    \left(x-\frac{L}{2}\right)^2+\left(y-\frac{L}{2}\right)^2\leq r_0^2.
\end{equation}
For each electron and positron within this region, the momentum components are transformed according to
\begin{equation}
    \begin{pmatrix}
        p_x'\\
        p_y'
    \end{pmatrix} =
    \begin{pmatrix}
        \cos\Delta\theta&-\sin\Delta\theta\\
        \sin\Delta\theta&\cos\Delta\theta
    \end{pmatrix}
    \begin{pmatrix}
        p_x\\
        p_y
    \end{pmatrix},\qquad p_z' = p_z.
\end{equation}
Here, $\Delta\theta$ is the imposed rotation angle, while $\boldsymbol{p}$ and $\boldsymbol{p}'$ denote the momenta before and after the perturbation, respectively. This perturbation preserves the magnitude of the particle momentum $|\boldsymbol{p}| = (p_x^2+p_y^2+p_z^2)^{1/2}$ and therefore its kinetic energy. The fiducial rotation is $\Delta\theta = \pi$, corresponding to a complete reversal of the in-plane momentum within the perturbed region. 

We performed a large number of paired simulations to scan across parameters associated with the initial perturbation and the global plasma properties. Each experiment compares an unperturbed baseline simulation (Sec.~\ref{sec:phys_sys_config}) with an otherwise identical perturbed counterpart. We will organize the parameter space into the categories of perturbation size, perturbation angle, and plasma physical parameters.

\subsection{Perturbation size and strength}

To examine the dependence of our results on the spatial extent of the initial perturbation, we considered
\[
r_0\in
\left\{
\frac{L}{512}, \,
\frac{L}{1024}, \,
\frac{L}{2048}, \,
\frac{L}{4096}, \,
\frac{L}{8192}
\right\},
\]
with $\Delta \theta = \pi$ in the sub-relativistic temperature regime. Note that three of these values correspond to a radius smaller than one grid-cell width ($\Delta x = L/1536$). We present results for $r_0 = L/512, L/2048$ and $L/8192$, which span the range of simulated perturbation sizes. The smallest size, $r_0 = L/8192$, is such that only a couple of simulated particles per species are expected to be within the perturbed region (noting that $N_{\rm ppc} \pi r_0^2/(\Delta x \Delta y) = 9\pi/8$). The purpose of this scan is to determine whether progressively more localized perturbations would still exhibit measurable growth and eventual decorrelation over the duration of the simulation.

After reaching a perturbation radius of $r_0 = L/8192$, we held $r_0$ fixed and instead decreased the rotation angle $\Delta\theta$. Our preliminary tests decreased the angle from $\Delta\theta=\pi$ by successive factors of four. In this paper, we only show results from a more extreme range,
\[
\Delta\theta\in
\left\{
\pi,
\frac{\pi}{10^5},
\frac{\pi}{10^{10}},
\frac{\pi}{10^{15}}
\right\}.
\]

\subsection{Sub-relativistic and ultra-relativistic temperature regimes \label{theta_i}}

In order to accurately characterize how the perturbation evolution depends on plasma temperature, we performed simulations in both sub-relativistic ($T_0 \ll m_e c^2$) and ultra-relativistic ($T_0 \gg m_e c^2$) temperature regimes. The sub-relativistic simulations used 
\[
\frac{T_0}{m_e c^2}\in
\{0.1,0.05,0.025,0.0125,0.00625\},
\]
whereas the ultra-relativistic simulations used $T_0/m_e c^2=10$. The sub-relativistic simulations were evolved for a maximum duration of $46.0 L/c$ (in our most extreme case of $T_0/m_e c^2 = 0.00625$, this is equivalent to $\sim5.14L/c_s$), and the ultra-relativistic simulations were evolved for $14.7 L/c$. The longer duration of the sub-relativistic cases is necessary to capture the slower perturbation evolution and eventual saturation. Some simulations did not reach these maximum durations. We therefore choose reference saturation times separately in Sec.~\ref{sec:definitions}, based on values available for all relevant simulations within each regime.

\subsection{Variation of plasma beta}
To investigate the sensitivity to a perturbation for plasmas with different magnetization, we performed a scan across different values of $\beta=8\pi n_0 T_0/B_0^2$. We tested three values of $\beta$ in both sub-relativistic ($T_0/m_e c^2 = 0.05$) and ultra-relativistic regimes, with the fiducial value being $\beta=1$, but we also explored $\beta =0.25$ and $\beta =  4$. This range is sufficient to infer the general trend of our results with $\beta$, while avoiding numerical challenges arising from extreme values of $\beta$.

\subsection{\label{sec:definitions}Difference diagnostics and normalization}
To characterize the growth of the perturbation over time, we consider two quantities. First, we define the root-mean-square difference in electric current density in a similar way to Eq.~\ref{jrms_def_new},
\begin{equation}
|\Delta\boldsymbol{J}|_{\rm rms}=\sqrt{\langle|\boldsymbol{J}_1 - \boldsymbol{J}_2|^2\rangle} \, , \label{jrms_def}
\end{equation}
where $\langle \cdot \rangle$ denotes an average over the simulation domain and $\boldsymbol{J}_1$ and $\boldsymbol{J}_2$ are the total current densities from each of the two (baseline and perturbed) simulations. The total current density in each simulation is $\boldsymbol{J}=\boldsymbol{J}_-+\boldsymbol{J}_+$, which accounts for contributions from both the electrons and positrons  ($\boldsymbol{J}_\mp$).

It is beneficial to normalize $|\Delta \boldsymbol{J}|_{\rm rms}$ to obtain a dimensionless quantity, though there is not a clear choice for doing so in terms of initial physical parameters. Instead, we define a characteristic current density scale, $J_0$, and normalize the difference as $|\Delta\boldsymbol{J}|_{\rm rms}/J_0$, where
\begin{equation}
J_0=\langle|\boldsymbol{J}(t_f)|^2\rangle^{1/2}. \label{eqn:J_0}
\end{equation}
Here, $\boldsymbol{J}$ is the current density of the baseline simulation, and $t_f$ is a fixed late time selected after the paired simulations have decorrelated. We use $t_fc/L\approx30.0$ for the sub-relativistic simulations and $t_fc/L\approx6.45$ for the ultra-relativistic simulations.

For the second diagnostic, we define the root-mean-square difference between the magnetic fields of the baseline and perturbed simulations
 \begin{equation}
|\Delta\boldsymbol{B}|_{\rm rms}=\sqrt{\langle|\boldsymbol{B}_1 - \boldsymbol{B}_2|^2\rangle} \, , \label{eqn:dbrms}
\end{equation}
which we normalize as $|\Delta\boldsymbol{B}|_{\rm rms}/B_0$, where the magnitude $B_0$ is defined in Sec.~\ref{sec:phys_sys_config}.

\subsection{Numerical noise and numerical precision floor\label{noise}}

As is well known in the community, PIC simulations are affected by discrete shot noise due to the limited number of macroparticles representing the plasma distribution function\citep{Touati_2022,EVSTATIEV2021110394} (e.g., 32 particles per cell for our simulations). This noise is an unavoidable consequence of the PIC methodology and does not have a direct physical origin (though it bears some similarities to thermal fluctuations). The noise will always be present, but can be reduced by increasing the number of macroparticles in the simulation\citep{Touati_2022,EVSTATIEV2021110394}.

PIC noise may cause the initial perturbation to grow (either on a light-crossing or sound-crossing timescale, depending on the physical regime). This numerical growth could be faster than the initial dynamical growth of the perturbation. 

In addition, floating-point precision imposes a hard limit on the smallest perturbation that can be resolved in simulations. In our analysis, we use double precision. A quantity such as $|\Delta \boldsymbol{J}|_{\rm rms}$ may fall below the floating-point precision even if the simulation is able to internally resolve differences. This causes an artificial floor to observable quantities which we call the "numerical precision floor".

\section{Results in the sub-relativistic regime} \label{sec:results}
We begin with the sub-relativistic case, using $T_0/m_ec^2=0.05$ and $\beta=1$ as our fiducial simulation pair. In this fiducial pair, the perturbed simulation differs from the corresponding baseline simulation only through an initial momentum rotation of $\Delta\theta = \pi$ applied within a circular region of radius $r_0=L/8192$. In later subsections, all of these reference parameters will be varied individually. Despite the localized initial perturbation being small, the resulting difference between the pair of simulations is found to spread and grow to produce macroscopically distinct, decorrelated plasma states. We characterize this evolution using the two primary diagnostics defined in Sec.~\ref{sec:definitions}, those diagnostics being the current density difference $\Delta\boldsymbol{J}$ and the magnetic field difference $\Delta\boldsymbol{B}$. To differentiate between the contributions from turbulence and from noise, we also analyze two control pairs: one with external driving alone turned off and another with both external driving and initial magnetic field fluctuations turned off.

\subsection{Visual growth of current density difference \label{current_density_difference_section}}

\begin{figure*}
    \centering
    \includegraphics[width=1\linewidth]{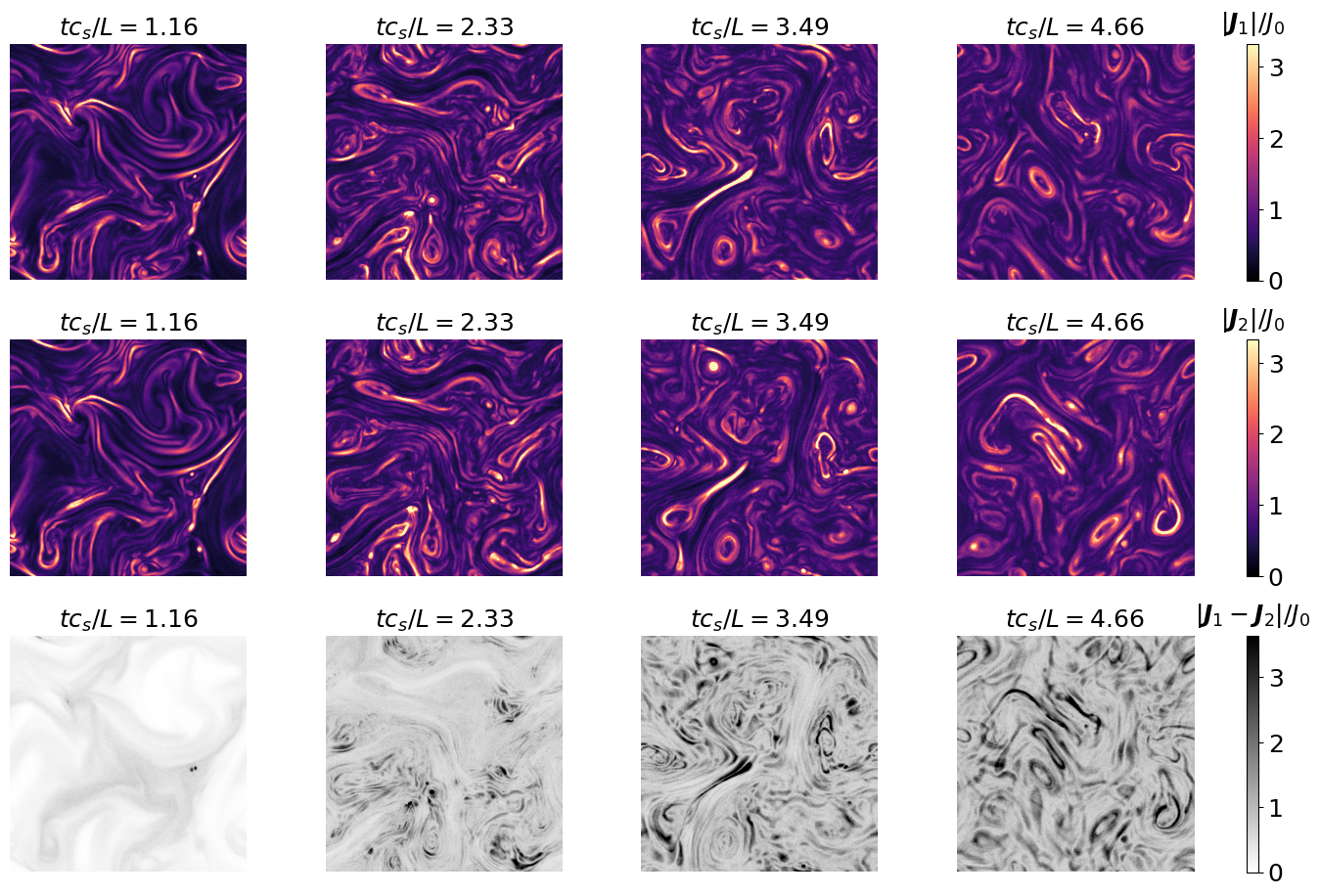}
    \caption{Evolution of the electric current density in the fiducial sub-relativistic simulation with $T_0/m_ec^2=0.05$ and $\beta=1$. From top to bottom, the rows show the current density magnitude in the unperturbed (baseline) simulation, $|\boldsymbol{J}_1|$; the current density magnitude in the perturbed simulation, $|\boldsymbol{J}_2|$; and the magnitude of their difference, $|\Delta\boldsymbol{J}|=|\boldsymbol{J}_1-\boldsymbol{J}_2|$. The columns show different numbers of sound-crossing times $tc_s/L$, as indicated above the panels. The perturbed simulation was initialized by rotating the particle momenta through $\Delta\theta=\pi$ within a circular region of radius $r_0=L/8192$. All current densities are normalized by the characteristic current density $J_0$ defined in Sec.~\ref{sec:definitions}. The plots of $|\boldsymbol{J}_1|$ and $|\boldsymbol{J}_2|$ share a common linear color scale, which differs from the linear scale of the plots of $|\Delta\boldsymbol{J}|$. In each scale, the lower limit is zero and the upper limit is the $99.5$th percentile of all values displayed in the corresponding rows; values above this limit are saturated.}
    \label{current_density_visual}
\end{figure*}

To characterize the divergence between baseline and perturbed simulations, we first examine the current density $\boldsymbol{J}$ for the fiducial sub-relativistic case. Fig.~\ref{current_density_visual} visually compares the baseline and perturbed simulations, along with the magnitude of their difference $|\Delta\boldsymbol{J}|$. At the earliest displayed time ($tc_s/L = 1.16$), both the baseline and the perturbed simulations appear visually indistinguishable from each other. As time elapses, the difference increases, resulting in the visually decorrelated pair seen at the latest time shown ($tc_s/L = 4.66$). The paired simulations therefore develop macroscopically distinct current density configurations from an initially localized perturbation, over the timescale of a few sound-crossing times.

\begin{figure*}
    \centering
    \includegraphics[width=1\linewidth]{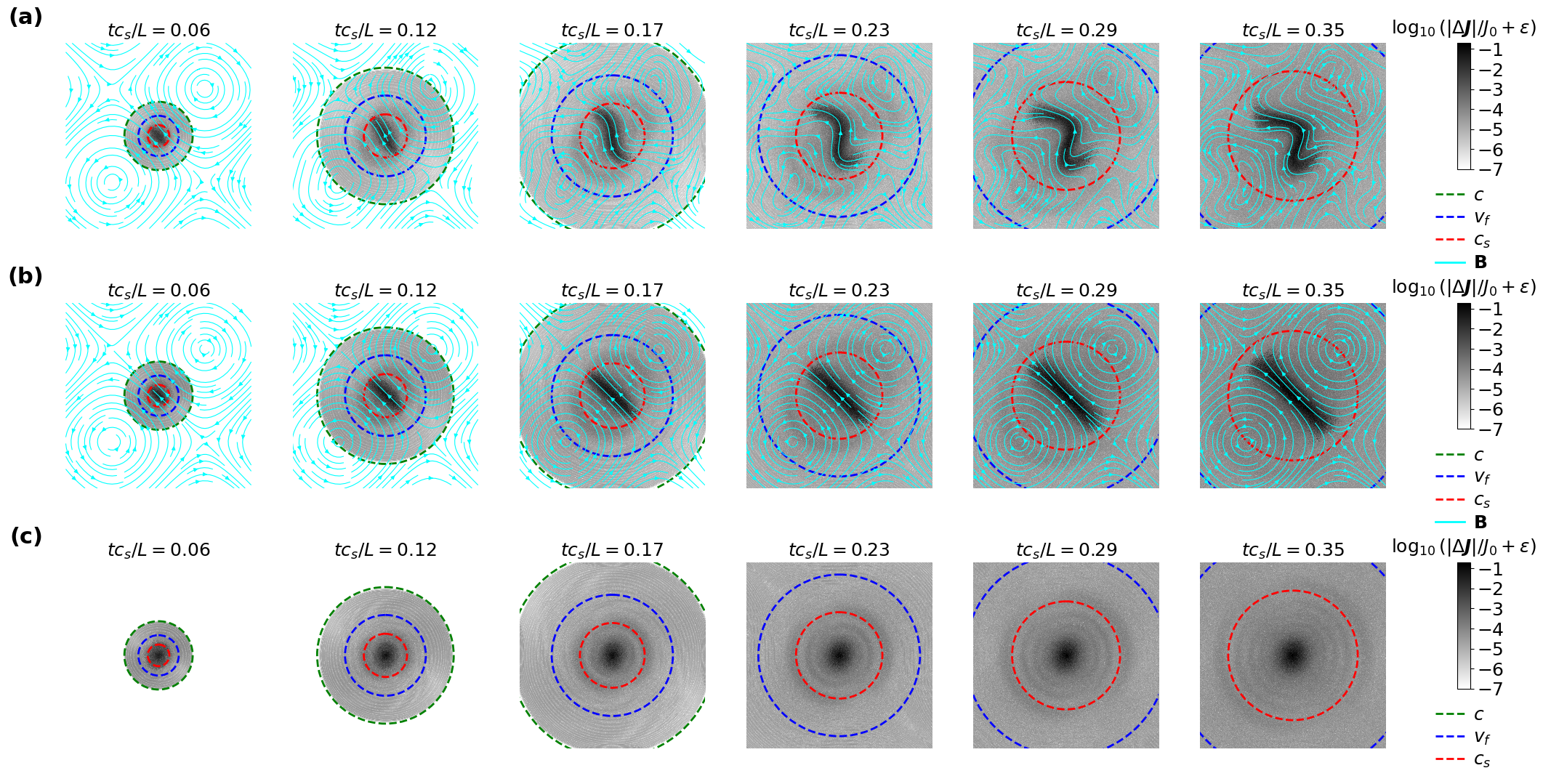}
    \caption{Early-time evolution of the current density difference for the same fiducial sub-relativistic simulation pair as in Fig.~\ref{current_density_visual} and two corresponding control pairs. The grayscale shows $\log_{10}(|\Delta\boldsymbol{J}|/J_0+\epsilon)$, where $|\Delta\boldsymbol{J}|=|\boldsymbol{J}_1-\boldsymbol{J}_2|$ and $\epsilon = 10^{-12}$ is a small parameter solely included to prevent taking the logarithm of zero. All three panels use the same grayscale range, allowing their displayed amplitudes to be compared directly. Panel (a) displays the fiducial baseline-perturbed simulation pair, which includes external driving and initial magnetic field fluctuations. Panel (b) displays the corresponding baseline-perturbed simulation pair with external driving turned off, but initial magnetic field fluctuations retained. Panel (c) displays the baseline-perturbed simulation pair with both external driving and the initial magnetic field fluctuations turned off. These controls help assess the contribution of numerical differences to the measured growth. The dashed circles are centered on the initial perturbation and show the propagation radii $r(t) = ct$ (green), $r(t) = v_ft$ (blue), and $r(t)=c_st$ (red). The cyan streamlines in panel (a) represent the in-plane magnetic field $(B_x,B_y)$ from the fiducial baseline simulation. The times $tc_s/L$ are indicated above the panels.}
    \label{early_current_density_visual}
\end{figure*}

To characterize the initial spatial growth of the perturbation, we next examine $|\Delta\boldsymbol{J}|$ at substantially earlier times. In Fig.~\ref{early_current_density_visual}, $|\Delta\boldsymbol{J}|$ is displayed on a logarithmic scale and compared with the propagation distances of the speed of light, $c$; the fast magnetosonic speed, $v_f$; and the sound speed, $c_s$. Fig.~\ref{early_current_density_visual} (a) shows that $|\Delta \boldsymbol{J}|$ develops primarily within the sound speed radius $r_s = c_s t$. A substantially weaker difference is visible outside of this region, bounded by the information front $r_c = c t$. Because its amplitude is several orders of magnitude below that of the central structure, this outer, low-amplitude portion of the perturbation contributes substantially less to the overall rms difference $|\Delta\boldsymbol{J}|_{\rm rms}$. The central $|\Delta \boldsymbol{J}|$ structure develops a 'snake-like' geometry aligned with the local in-plane magnetic field structure (associated with the initial conditions in Eq.~\ref{eq:ic}). This correspondence suggests that the local magnetic geometry influences the spatial development of the perturbation; in particular, the perturbation spreads preferentially along the magnetic field lines rather than across them. Such an anisotropy is expected because particles stream along magnetic field lines unimpeded, while they do not easily cross field lines due to Larmor gyration; thus the thermal transport of information occurs preferentially along the field lines.

Figs.~\ref{early_current_density_visual} (b) and (c) help isolate the numerical contribution to perturbation growth by using two control pairs. In panel (b), external driving is turned off while the initial magnetic field fluctuations are retained. In panel (c), both the external driving and initial magnetic field fluctuations are turned off, such that $\boldsymbol{B}=B_0\hat{\boldsymbol{z}}$ rather than the configuration given by Eq.~\ref{eq:ic}. In contrast to panel (a), the high-amplitude central region exhibits little spatial growth for the case in panel (c), while a much weaker, approximately circular signal still spreads outwards near the information front for all cases. The presence of this weak and rapidly expanding signal in both of the control cases indicates that numerical noise contributes to $|\Delta\boldsymbol{J}|$ outside of the slowly expanding central region. It should be noted that even at the earliest displayed time ($tc_s/L = 0.06)$, the high-amplitude region is much larger than the initial scale of the perturbation ($r_0 = L/8192$), and may be comparable to a few characteristic gyroradii ($\rho_{e0} = L/128$). This is especially notable in Fig.~\ref{early_current_density_visual} (c), where the absence of both external driving and initial magnetic field fluctuations isolates the contribution of numerical noise. At later times, not shown in panels (b) and (c), the low-amplitude difference spreads throughout the domain, and the rms difference eventually saturates, whereas the high-amplitude central region continues to expand slowly. Later snapshots are omitted to preserve a direct comparison with panel (a) and to show the early spatial structure in more detail. 

Comparing panels (b) and (c) illustrates the impact of the initial magnetic field fluctuations on the perturbation's overall shape. In panel (b), the elongated central structure remains aligned with the in-plane magnetic geometry despite the absence of external driving. When the initial magnetic fluctuations are also removed in panel (c), $|\Delta\boldsymbol{J}|$ instead develops a more circular structure. This confirms that the initial magnetic field helps shape the perturbation.

\subsection{Quantitative growth of current density difference\label{quant_current_diff}}
\begin{figure}
    \centering
    \includegraphics[width=1\linewidth]{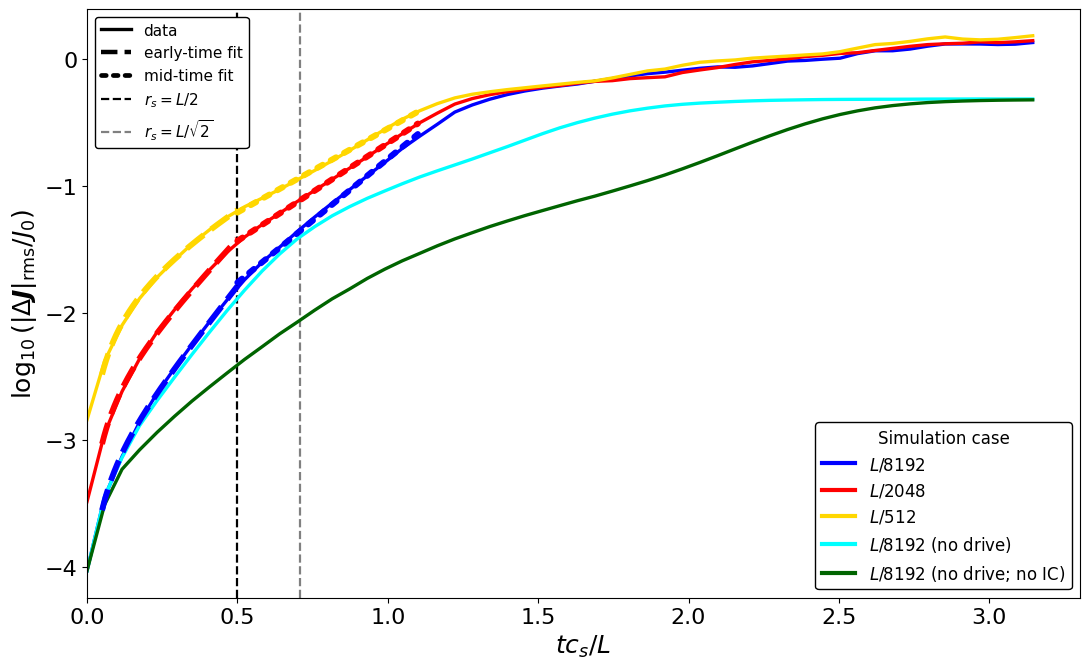}
    \caption{Evolution of the normalized, logarithmic rms total current density difference, $\log_{10}(|\Delta\boldsymbol{J}|_{\rm rms}/J_0)$, in the sub-relativistic simulations with $T_0/m_ec^2=0.05$, $\beta=1$, and $\Delta\theta=\pi$. The colors distinguish the initial perturbation radii $r_0=L/8192$ (blue), $r_0=L/2048$ (red), and $r_0 = L/512$ (yellow). The cyan line shows the $r_0=L/8192$ case but with external driving turned off. The green line shows the $r_0=L/8192$ case with both external driving and initial magnetic field fluctuations turned off. Solid curves show the simulation data. Dashed curves show fits to the early-time model from Sec.~\ref{early_time_pert} over $0.05\leq tc_s/L\leq0.5$, while dotted curves show empirical exponential fits over $0.5\leq tc_s/L\leq1.1$. Fits are shown only for the three driven cases and use the same color as their corresponding simulation curves. The vertical black and gray dashed lines mark the times at which the sound front, $r_s(t)=c_st$, reaches $r_s=L/2$ and $r_s=L/\sqrt{2}$, respectively.}
    \label{subrel_jsrmsdiff}
\end{figure}

To quantify the divergence between the baseline and perturbed simulations, we compute the root-mean-square current density difference, $|\Delta\boldsymbol{J}|_{\rm rms}$, defined in Eq.~\ref{jrms_def}. Fig.~\ref{subrel_jsrmsdiff} shows the normalized difference, $|\Delta\boldsymbol{J}|_{\rm rms}/J_0$, as a function of the number of sound-crossing times, $tc_s/L$. Although our discussion focuses on the fiducial perturbation radius of $r_0=L/8192$, we also include the $L/2048$ and $L/512$ cases to show the dependence on the initial spatial size of the perturbation. We also include the two control cases shown in Fig.~\ref{early_current_density_visual}, both of which use $r_0 = L/8192$.

At early times, the $r_0=L/8192$ case is well described by the early-time model derived in Sec.~\ref{early_time_pert}. Fitting over $0.05\leq tc_s/L\leq0.5$ gives $J_{\rm char}/J_0=5.00\times10^{-4}$ and $\lambda=6.98c_s/L$. The upper limit of the fitting interval corresponds to $r_s=c_st=L/2$, when the circular sound front first reaches the nearest boundary of the simulation domain. Beyond this point, the expanding circular geometry assumed in Eq.~\ref{eq:fit}
is no longer applicable.

Applying the same fitting model and interval independently to the larger perturbations gives $J_{\rm char}/J_0=2.01\times10^{-3}$ and $\lambda=5.92c_s/L$ for $r_0=L/2048$, and $J_{\rm char}/J_0=1.30\times10^{-2}$ and $\lambda=3.38c_s/L$ for $r_0=L/512$. As expected, $J_{\rm char}/J_0$ increases with $r_0$ because the initial perturbation affects a larger region and therefore produces a larger rms current density difference. Although the larger perturbations consequently exhibit faster absolute growth at the earliest times, their fitted values of $\lambda$ are smaller. 

After the propagation front reaches the nearest periodic boundary (this transition is described in Sec.~\ref{int_time_evolution}), the circular geometry assumed by the early-time model no longer applies. However, $|\Delta\boldsymbol{J}|_{\rm rms}$ continues to grow approximately exponentially over a separate intermediate interval. We describe this intermediate-time stage empirically using the fit
\begin{equation}
    \frac{|\Delta\boldsymbol{J}|_{\rm rms}}{J_0}
    = Ce^{\lambda_{\rm mid}t}.\label{eq:current_mid_fit}
\end{equation}
For the $L/8192$ case, fitting over $0.5\leq tc_s/L\leq1.10$ gives $C=1.84\times10^{-3}$ and $\lambda_{\rm mid} =4.49c_s/L$. The $L/2048$ case gives $C=6.29\times10^{-3}$ and $\lambda_{\rm mid}=3.55c_s/L$, while the $L/512$ case gives $C=1.35\times10^{-2}$ and $\lambda_{\rm mid}=3.05c_s/L$ over the same region. The close agreement with a simple exponential (which is not a continuation of the early-time model) over this interval is an empirical observation, and we do not presently provide a physical model for this stage.

Subsequently, the three driven curves converge shortly after $tc_s/L\sim1$ and follow nearly identical trajectories toward similar late-time amplitudes. Beyond the range shown in Fig.~\ref{subrel_jsrmsdiff}, all three cases saturate near $|\Delta\boldsymbol{J}|_{\rm rms}/J_0\approx\sqrt{2}$, as expected for statistically decorrelated current density fields (Sec.~\ref{latetimemodel}). Thus, within the range of initial radii considered here, the perturbation size strongly affects the early rms difference but has little effect on the eventual late-time outcome.

We also show the two undriven control cases in Fig.~\ref{subrel_jsrmsdiff}: one with initial magnetic fluctuations (cyan) and one without initial magnetic fluctuations (green). The cyan curve corresponds to the spatial evolution shown in Fig.~\ref{early_current_density_visual} (b), while the green curve corresponds to Fig.~\ref{early_current_density_visual} (c). Both of these curves help to assess the effect of numerical noise on the measured divergence. Compared with the driven fiducial case, the control case with no initial magnetic field fluctuations exhibits slower growth over most of the displayed interval and approaches a substantially lower saturation plateau. In contrast with this, the control case with initial magnetic field fluctuations follows a path nearly identical to that of the corresponding driven case for much of the early-time expansion, before approaching a similar plateau to the other undriven case. This difference demonstrates that the initialized magnetic field fluctuations strongly influence the early amplification of the perturbation, even in the absence of external driving.

All curves in Fig.~\ref{subrel_jsrmsdiff} are normalized by the driven baseline simulation's value of $J_0$, allowing the absolute rms differences in the driven and undriven cases to be compared using a common current scale. Under this normalization, both undriven cases approach a late-time saturation level approximately $2.9$ times smaller than that of the driven cases. The characteristic current density of the undriven simulations is also approximately $2.9$ times smaller, indicating that the lower level of saturation remains consistent with the statistical decorrelation of the undriven systems. Thus, the control cases develop a smaller absolute current density difference than the driven simulations, while still suggesting that numerical noise may amplify the perturbation sufficiently to contribute to eventual decorrelation, even in the absence of external driving or initial magnetic field fluctuations. 

\subsection{Magnetic field evolution \label{mag_field}}

The observed divergence of the current density between the simulation pairs motivates examining whether similar behavior develops in other fields. As an example, we now describe a similar analysis using the magnetic field $\boldsymbol{B}$. For this, we consider the magnetic field difference $\Delta\boldsymbol{B}=\boldsymbol{B}_1-\boldsymbol{B}_2$ described in Sec.~\ref{sec:definitions}. Note that $\boldsymbol{B}$ is related to $\boldsymbol{J}$ through the Amp\`{e}re-Maxwell equation, which gives $\boldsymbol{J} \approx c \nabla \times \boldsymbol{B}/4\pi$ when displacement current is neglected (as appropriate for turbulence in our simulations, since bulk motions are sub-relativistic). For typical (Kolmogorov-like) turbulence, $\boldsymbol{B}$ is dominated by large-scale structure (at the driving scale), while $\boldsymbol{J}$ is organized in intermittent, kinetic-scale current sheets.

\begin{figure*}
    \centering

    \includegraphics[width=1\linewidth]{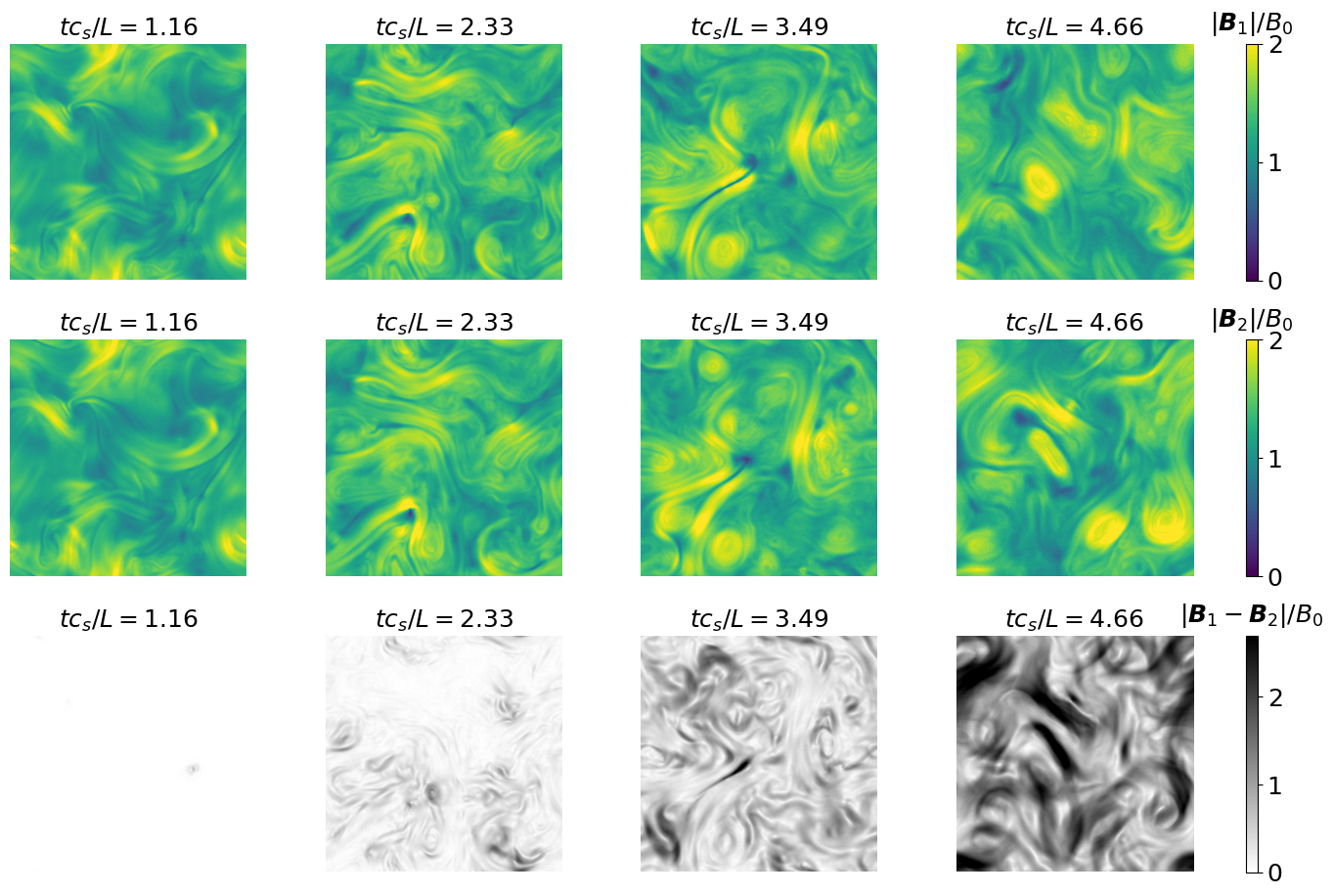}
    \caption{Evolution of the magnetic field in the fiducial sub-relativistic simulation with $T_0/m_ec^2=0.05$, $\beta=1$, $r_0=L/8192$, and $\Delta\theta=\pi$. From top to bottom, the rows show the magnetic field magnitude in the unperturbed (baseline) simulation, $|\boldsymbol{B}_1|$; the magnetic field magnitude in the perturbed simulation, $|\boldsymbol{B}_2|$; and the magnitude of their difference, $|\Delta\boldsymbol{B}|=|\boldsymbol{B}_1-\boldsymbol{B}_2|$. The columns show different numbers of sound-crossing times $tc_s/L$, as indicated above the panels. All quantities are normalized by the initial mean magnetic field $B_0$. The first two rows share a common linear color scale, while the difference row uses a separate linear scale. For each scale, the upper limit is the $99.5$th percentile of the values across the corresponding displayed panels, with larger values saturated.}
    \label{fig:subrel_mag_visual}
\end{figure*}

Fig.~\ref{fig:subrel_mag_visual} provides a visualization of the magnetic field in the fiducial sub-relativistic case. Both the baseline and perturbed simulations ($\boldsymbol{B}_1$ and $\boldsymbol{B}_2$) remain visually similar at the earliest displayed time, with noticeable visual differences starting to appear at $tc_s/L=2.33$, and the two appear visually distinct throughout the domain by $tc_s/L=4.66$. Their difference, $|\Delta\boldsymbol{B}|$, further illustrates this evolution: a small localized difference at the earliest displayed time grows to encompass the entire spatial domain by the latest displayed time. 

\begin{figure*}
    \centering
    \includegraphics[width=1\linewidth]{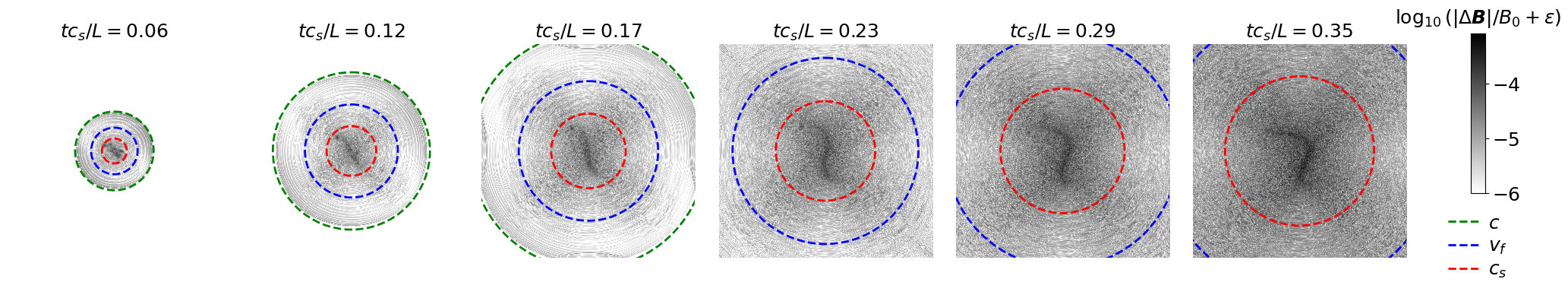}
    \caption{Early-time evolution of the magnetic field difference for the same fiducial sub-relativistic case as in Fig.~\ref{fig:subrel_mag_visual}. The grayscale shows $\log_{10}(|\Delta\boldsymbol{B}|/B_0+\epsilon)$, where $|\Delta\boldsymbol{B}|=|\boldsymbol{B}_1-\boldsymbol{B}_2|$ and $\epsilon = 10^{-12}$ is a small parameter solely included to prevent taking the logarithm of zero. The upper color-scale limit is the maximum value across displayed snapshots, while values below $-6$ are clipped. The dashed circles are centered on the initial perturbation and show the propagation radii $r(t) = ct$ (green), $r(t) = v_ft$ (blue), and $r(t)=c_st$ (red). The times $tc_s/L$ are indicated above the panels.}   \label{fig:subrel_mag_early}
\end{figure*}

The early evolution of $|\Delta\boldsymbol{B}|$ is examined more closely in Fig.~\ref{fig:subrel_mag_early}. The early-time structure is notably similar in geometry to that of $|\Delta\boldsymbol{J}|$ in Fig.~\ref{early_current_density_visual}. The primary visual difference is that $|\Delta\boldsymbol{B}|$ appears considerably grainier than $|\Delta\boldsymbol{J}|$. This may reflect a larger relative contribution from the numerical noise because the magnetic field response lies closer to the background amplitude. The difference in amplitudes between $|\Delta\boldsymbol{B}|$ and $|\Delta\boldsymbol{J}|$ may result from the weaker initial magnetic field response to the perturbation, since the imposed perturbation directly changes the particle momenta and hence the current density, whereas the magnetic field responds subsequently through the Amp\`{e}re-Maxwell equation. A low-amplitude X-shaped structure (aligned with the cell diagonals) is also visible within the information front in Fig.~\ref{fig:subrel_mag_early} and may be associated with this numerical background. 

\begin{figure}
    \centering
    \includegraphics[width=1\linewidth]{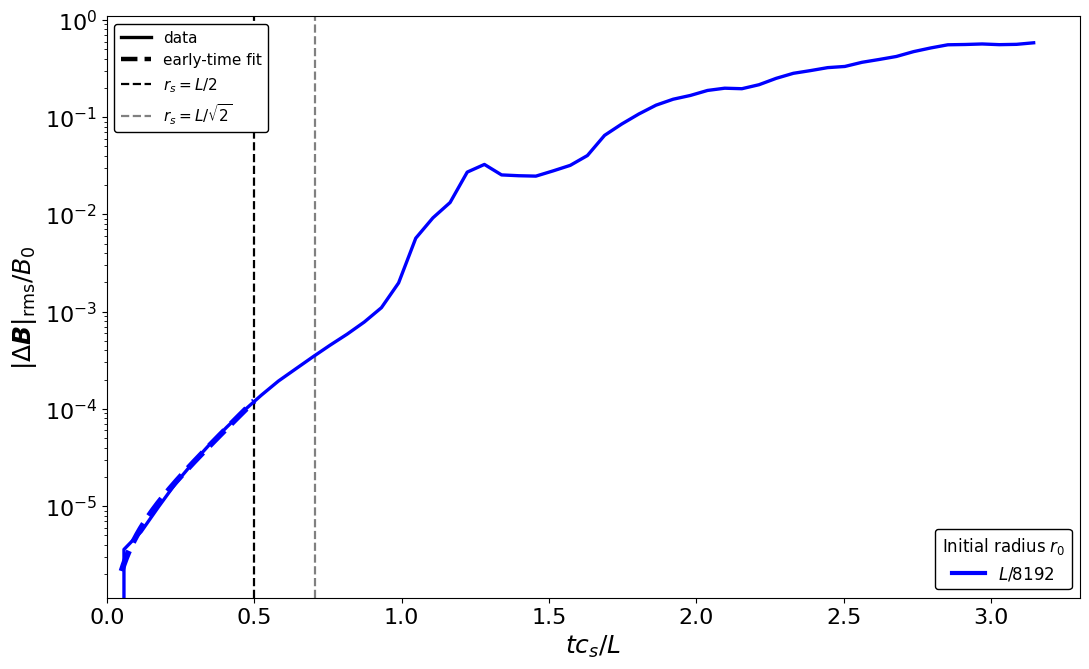}
    \caption{Evolution of the normalized rms magnetic field difference, $|\Delta\boldsymbol{B}|_{\rm rms}/B_0$, in the fiducial sub-relativistic simulation with $T_0/m_ec^2=0.05$, $\beta=1$, $\Delta\theta=\pi$, and $r_0=L/8192$. The solid curve shows the simulation data, and the dashed curve shows a fit to the early-time model from Sec.~\ref{early_time_pert} over $0.05\leq tc_s/L\leq0.5$. The vertical black and gray dashed lines mark the times at which the sound front, $r_s(t)=c_st$, reaches $r_s=L/2$ and $r_s=L/\sqrt{2}$, respectively.}
    \label{subrel_magrmsdiff}
\end{figure}

We next compute the normalized rms magnetic field difference $|\Delta\boldsymbol{B}|_{\rm rms}/B_0$, defined in Sec.~\ref{sec:definitions}. In Fig.~\ref{subrel_magrmsdiff}, the evolution of $|\Delta\boldsymbol{B}|_{\rm rms}/B_0$ displays the same broad trends as observed with $|\Delta\boldsymbol{J}|_{\rm rms}/J_0$ in Sec.~\ref{quant_current_diff}, showing an initial period of rapid growth, followed by a transition towards decorrelation. Applying the model from Sec.~\ref{early_time_pert} over the interval $0.05\leq tc_s/L\leq0.5$ gives a growth rate of $\lambda_B=6.87c_s/L$ and an amplitude $B_{\rm char}/B_0 = 3.91 \times 10^{-6}$. This is very close to the value of $\lambda_J=6.98c_s/L$ (from Sec.~\ref{quant_current_diff}) for the total current density difference, indicating that the current density and magnetic field differences undergo comparable early-time exponential amplification. By comparison, the fitted amplitude of the current density difference is much larger ($J_{\rm char}/J_0 = 5.00\times10^{-4}$); note that the fitted amplitudes should not be expected to agree, as the two quantities use different normalizations and respond differently to the initial perturbation.

Nevertheless, $|\Delta\boldsymbol{B}|_{\rm rms}/B_0$ is less accurately fitted by the model than $|\Delta\boldsymbol{J}|_{\rm rms}/J_0$, partly because it exhibits a sizable "bump" that begins near $tc_s/L \sim 1$. This irregularity may arise because the magnetic field varies on larger scales than the current density and thus is more susceptible to statistical fluctuations (associated with driving and turbulence). Because this intermediate interval cannot be represented reliably by a single exponential model, and the same feature appears in the other simulations, we do not report a separate intermediate-time fit for the magnetic field. The early-time agreement should therefore be interpreted as evidence that the same model is broadly applicable to both $|\Delta\boldsymbol{B}|_{\rm rms}$ and $|\Delta\boldsymbol{J}|_{\rm rms}$.

The parameter dependence of the magnetic field evolution was explored in a similar way to the current density, by varying the initial perturbation radius, rotation angle, and plasma temperature. These scans reproduce similar qualitative dependencies similar to those of the current density difference (which will be described in the following subsections), and reveal no additional unique behavior. Therefore, we do not show the magnetic field results separately for these additional scans.

\subsection{Dependence on perturbation angle}

We next examine the dependence on the strength of the initial perturbation by varying the momentum's angle of rotation $\Delta\theta$ while holding the perturbation radius fixed at $r_0=L/8192$. As described in Sec.~\ref{angular_dependence_model}, the initial current density difference satisfies $|\Delta\boldsymbol{J}|\propto\Delta\theta$ for small $\Delta\theta$ (see Eq.~\ref{eq:jchartheta}). The early-time model therefore predicts that changing $\Delta\theta$ should primarily affect the initial amplitude rather than the subsequent time dependence, provided the imposed perturbation remains above the numerical precision floor.

\begin{figure}
    \centering
    \includegraphics[width=1\linewidth]{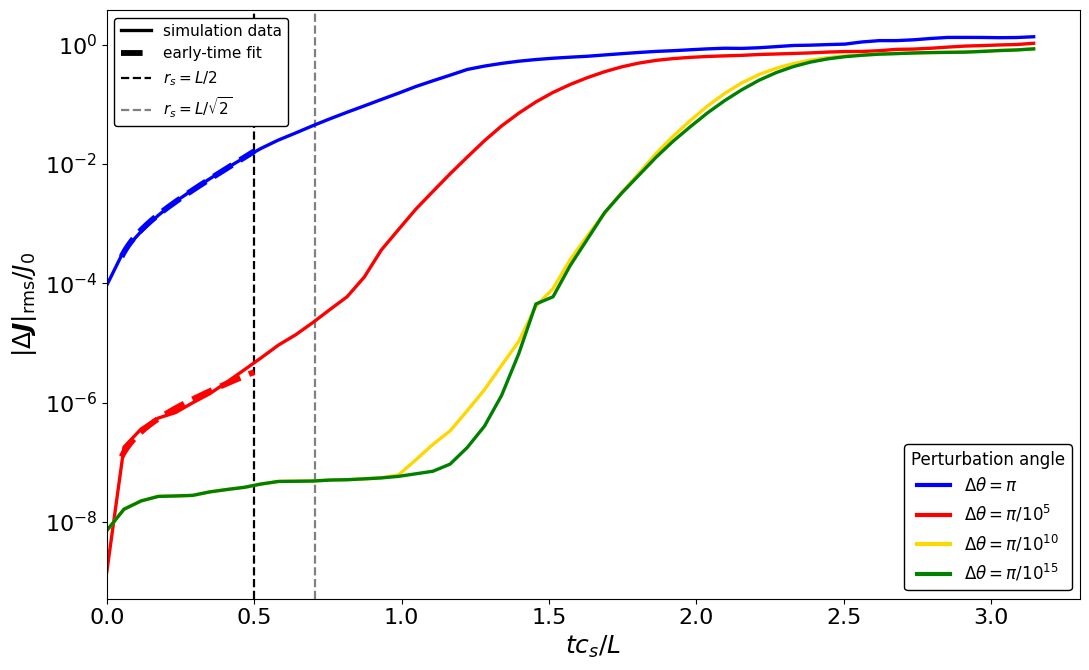}
    \caption{Dependence of the normalized rms total current density difference, $|\Delta\boldsymbol{J}|_{\rm rms}/J_0$, on the perturbation angle $\Delta\theta$ for the sub-relativistic simulations with $T_0/m_ec^2=0.05$, $\beta=1$, and $r_0=L/8192$. The four solid lines show simulation data for perturbation angles $\Delta\theta=\pi$ (blue), $\Delta\theta=\pi/10^5$ (red), $\Delta\theta=\pi/10^{10}$ (yellow), and $\Delta\theta=\pi/10^{15}$ (green). Dashed curves show fits of the early-time model from Sec.~\ref{early_time_pert} over the region $0.05\leq tc_s/L\leq 0.5$ for the $\Delta\theta=\pi$ and $\Delta\theta=\pi/10^5$ cases. Fits are not shown for the $\Delta\theta=\pi/10^{10}$ and $\Delta\theta=\pi/10^{15}$ cases because their early-time differences are dominated by a common numerical precision floor, and they do not follow the assumed early-time model. The vertical black and gray dashed lines mark the times at which the sound front, $r_s(t)=c_st$, reaches $L/2$ and $L/\sqrt{2}$, respectively.}
    \label{angles_graph}
    
\end{figure}

In Fig.~\ref{angles_graph}, we show the evolution of $|\Delta\boldsymbol{J}|_{\rm rms}/J_0$ for the scan in $\Delta\theta$. The $\Delta\theta=\pi$ and $\pi/10^5$ cases exhibit resolved early-time growth and can therefore be compared with the model from Sec.~\ref{early_time_pert}. As before, the fit is restricted to $0.05\leq tc_s/L\leq0.5$, before the sound front reaches the nearest domain boundary. Because $J_{\rm char}$ and $\lambda$ can be correlated within this model, the dependence on $\Delta\theta$ is more directly tested using the earliest stored values of $|\Delta\boldsymbol{J}|_{\rm rms}/J_0$ than by comparing $J_{\rm char}$ alone. The fits give $J_{\rm char}/J_0=3.65\times10^{-7}$ and $\lambda=4.44c_s/L$ for $\Delta\theta = \pi/10^5$, compared with $J_{\rm char}/J_0=5.00\times10^{-4}$ and $\lambda=6.98c_s/L$ for the fiducial $\Delta\theta = \pi$ case. Thus, the dependence predicted in Sec.~\ref{angular_dependence_model} is not fully reflected in the fitted parameters themselves. At  $t=0$, however, we obtain $|\Delta\boldsymbol{J}|_{\rm rms}/J_0=9.32\times10^{-5}$ for $\Delta\theta=\pi$ and $|\Delta\boldsymbol{J}|_{\rm rms}/J_0=1.46\times10^{-9}$ for $\Delta\theta=\pi/10^5$. Their ratio of $1.57\times10^{-5}$ is consistent with the difference of order $10^{-5}$ expected from the imposed rotation angles (Sec.~\ref{angular_dependence_model}).

If the imposed angular dependence remained numerically resolved, the initial differences for $\Delta\theta=\pi/10^{10}$ and $\Delta\theta = \pi/10^{15}$ would also differ by factors of $\sim 10^{-5}$. Instead, the two curves begin at nearly the same level and follow similar low-amplitude trajectories. This indicates that the imposed perturbations lie below the numerical precision floor (discussed in Sec.~\ref{noise}) and therefore cannot be distinguished from the numerical background. Their observable growth is therefore consistent with amplification of a common numerical background rather than a resolved dependence on the initial momentum rotation angle. Because the early-time model assumes that the measured difference represents the imposed perturbation, it consequently provides a poor fit to both cases. The two curves also exhibit a distinct uptick at approximately the same time ($t c_s/L \approx 1.5$) as the $\Delta\theta=\pi$ case begins to level off toward saturation. Although the imposed perturbations may initially grow beneath the numerical precision floor and become visible only after sufficient amplification, the present double-precision simulations cannot distinguish this possibility from amplification of the numerical background itself.

Despite the widely different initial perturbation angles, all four curves eventually converge and approach the same late-time saturation level $|\Delta\boldsymbol{J}|_{\rm rms}/J_0 \approx 2^{1/2}$ (at times later than shown in Fig.~\ref{angles_graph}). Thus, the perturbation angle controls the resolvable early-time difference, whereas the statistically decorrelated late-time state is largely independent of the initial perturbation.

\subsection{Dependence on initial temperature\label{temp}}

\begin{figure*}
    \centering
    \includegraphics[width=1\linewidth]{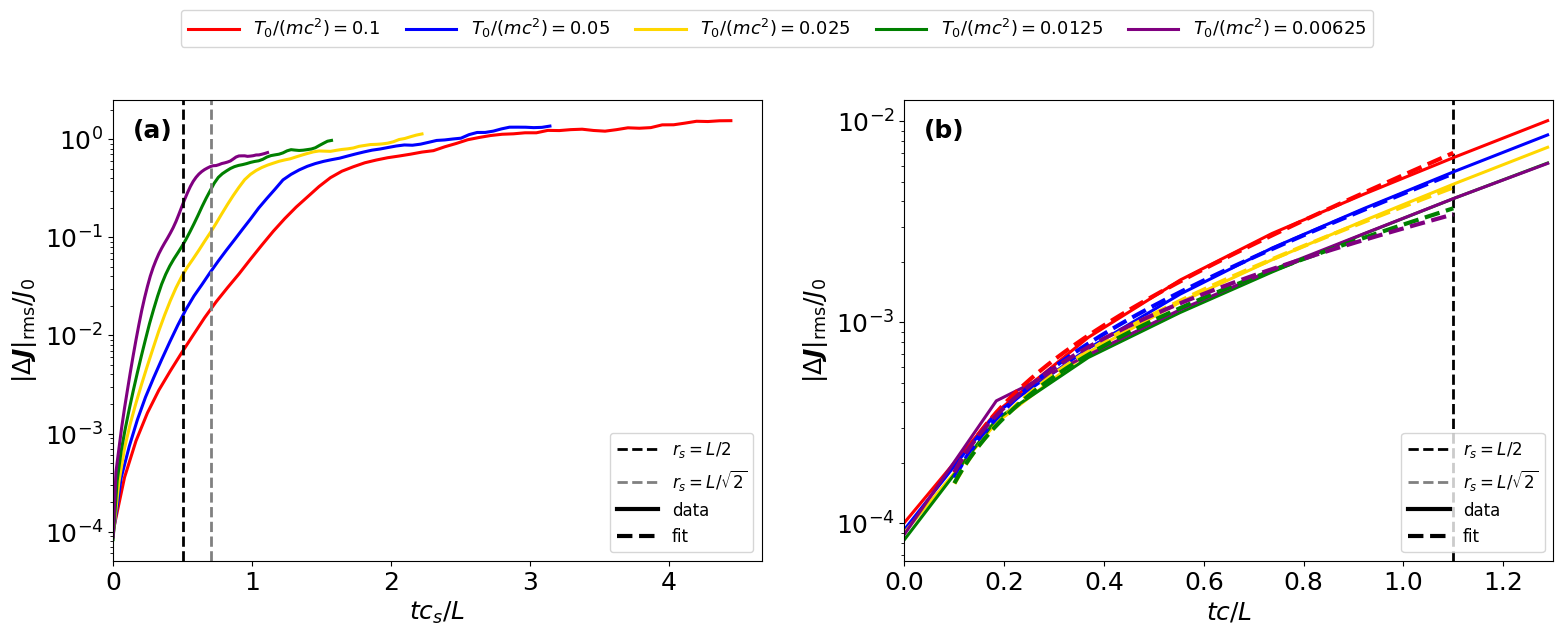}
    \caption{Temperature dependence of the evolution of the normalized rms current density difference, $|\Delta\boldsymbol{J}|_{\rm rms}/J_0$, for $T_0/m_ec^2\in\{0.1,0.05,0.025,0.0125,0.00625\}$. All simulations use $\beta=1$, $r_0=L/8192$, and $\Delta\theta=\pi$. Panel (a) shows the evolution as a function of the number of sound-crossing times $tc_s/L$, while panel (b) shows the same data as a function of the number of light-crossing times $tc/L$ (over a shorter interval, to emphasize the early-time differences). Dashed segments in panel (b) show fits performed over the interval $0.10\leq tc/L\leq1.30$. The vertical black and gray dashed lines in panel (a) indicate the times at which the sound front reaches $r_s=L/2$ and $r_s=L/\sqrt{2}$, respectively. The vertical black dashed line in panel (b) marks the upper boundary of the fitting interval.}
    \label{fig:temperature_scan}
\end{figure*}

We next examine the dependence of the divergence on the initial temperature within the sub-relativistic regime. We consider $T_0/m_ec^2\in\{0.1,0.05,0.025,0.0125,0.00625\}$ with $\beta=1$, $r_0=L/8192$, and $\Delta\theta=\pi$ as fixed parameters. Fig.~\ref{fig:temperature_scan} shows the resulting evolution of  $|\Delta\boldsymbol{J}|_{\rm rms}/J_0$ as a function of both the number of sound-crossing times $tc_s/L$ (panel a) and the number of light-crossing times $tc/L$ (panel b).

Fig.~\ref{fig:temperature_scan} illustrates that all five of the selected temperatures exhibit the same qualitative evolution with rapid early and intermediate-time growth followed by a slower leveling off to saturation at late times. The curves all approach a similar saturation level, confirming that the decorrelated state is insensitive to the initial temperature.

At a fixed value of $tc_s/L$, a lower temperature simulation has a smaller sound speed and thus has evolved for a larger number of light-crossing times, $tc/L$. Therefore, the growth that can be written as 
\begin{equation}
e^{\lambda t}
=
\exp\left[
\left(\frac{\lambda L}{c}\right)\frac{tc}{L}
\right]
=
\exp\left[
\left(\frac{\lambda L}{c_s}\right)\frac{tc_s}{L}
\right]
\end{equation}
has parameters satisfying
\begin{equation}
\frac{\lambda L}{c_s}
=
\frac{c}{c_s}\frac{\lambda L}{c}.
\end{equation}
The apparently faster growth at lower temperatures in panel (a) is therefore largely a consequence of the temperature-dependent time normalization.

To compare the simulations over the same physical interval, we perform the fits using light-crossing time, $tc/L$, and denote the resulting dimensional growth rate by $\lambda_c$. The fits shown in Fig.~\ref{fig:temperature_scan} (b) give $\lambda_c\in\{2.46c/L,2.20c/L,2.06c/L,1.65c/L,1.30c/L\}$ and $J_{\rm char}/J_0\in\{1.65\times10^{-3},1.56\times10^{-3},1.47\times10^{-3},1.50\times10^{-3},1.72\times10^{-3}\}$ for the respective temperatures $T_0/m_ec^2\in\{0.1,0.05,0.025,0.0125,0.00625\}$. The fitted value of $\lambda_c$ decreases by approximately $47\%$ from the highest to the lowest temperature, whereas $J_{\rm char}/J_0$ remains relatively constant. However, the agreement with the fitting function deteriorates for the lowest temperature, and we cannot justify assigning any particular functional dependence to the pattern given the limited parameter range. 

\subsection{Dependence on plasma beta}

\begin{figure*}
    \centering
    \includegraphics[width=1\linewidth]{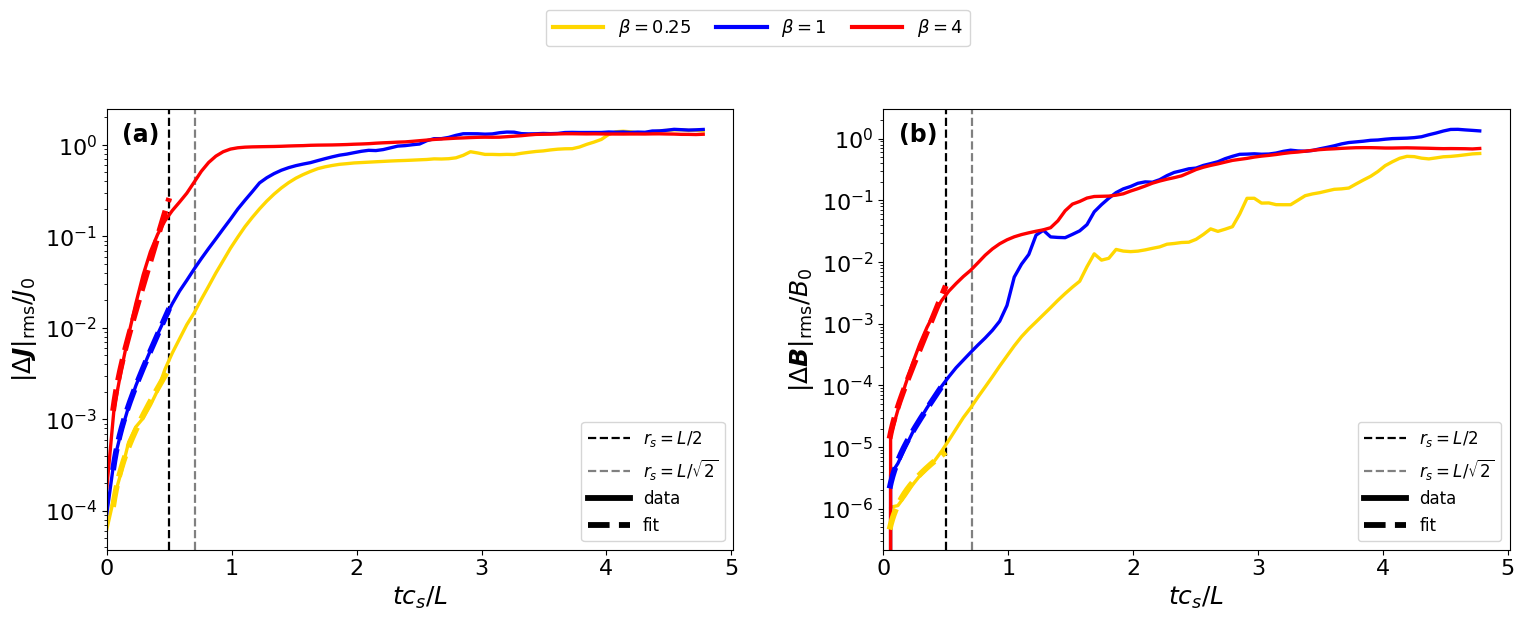}
    \caption{Dependence of the normalized rms current density and magnetic field differences on plasma beta. Panel (a) shows $|\Delta\boldsymbol{J}|_{\rm rms}/J_0$, and panel (b) shows $|\Delta\boldsymbol{B}|_{\rm rms}/B_0$, both as functions of the number of sound-crossing times, $tc_s/L$. The colors denote $\beta = 0.25$ (blue), $\beta = 1$ (red), and $\beta = 4$ (yellow). All simulations use $T_0/m_ec^2 = 0.05$, $r_0=L/8192$, and $\Delta\theta=\pi$. Solid curves show the simulation data, while the corresponding colored dashed segments show fits to the early-time model from Sec.~\ref{early_time_pert} over the interval $0.05\leq tc_s/L\leq0.5$. The vertical black and gray dashed lines indicate the times at which the sound front reaches $r_s=L/2$ and $r_s=L/\sqrt{2}$, respectively.}
    \label{fig:beta_dependence}
\end{figure*}
The growth rates of both the current density and magnetic field differences depend on plasma beta, $\beta$. Fig.~\ref{fig:beta_dependence} displays this, showing both the current density $|\Delta \boldsymbol{J}|_{\rm rms}/J_0$ (panel a) and the magnetic field $|\Delta\boldsymbol{B}|_{\rm rms}/B_0$ (panel b), with both exhibiting systematically different evolution across the three values of $\beta$. Comparing panels (a) and (b) to each other shows that each quantity retains its qualitative multi-stage evolution across all three values of $\beta$; more specifically, $|\Delta \boldsymbol{J}|_{\rm rms}/J_0$ displays relatively smooth, but rapid growth followed by saturation, while $|\Delta\boldsymbol{B}|_{\rm rms}/B_0$ shows a more gradual and irregular approach to saturation. Since $\boldsymbol{J} \approx (c/4\pi) \nabla \times \boldsymbol{B}$ when the displacement current is small, the current density is more sensitive to finer-scale magnetic structure. The faster growth of $|\Delta\boldsymbol{J}|_{\rm rms}/J_0$ therefore suggests that decorrelation develops first at smaller spatial scales and subsequently at larger scales, consistent with the more gradual growth of $|\Delta \boldsymbol{B}|_{\rm rms}/B_0$. Quantitatively, we find that as $\beta$ increases from $0.25$ to $4$, the fitted current density growth rate $\lambda_J$ rises from $5.19c_s/L$ to $10.50c_s/L$, while the fitted magnetic field growth rate $\lambda_B$ rises from $2.99c_s/L$ to $11.39c_s/L$. The close agreement between $\lambda_J$ and $\lambda_B$ found for the fiducial $\beta = 1$ case thus does not persist at $\beta = 0.25$, where the fitted current density growth rate is substantially larger than the magnetic field growth rate. At higher $\beta$, a larger fraction of the plasma's energy is associated with particle thermal motions relative to the magnetic field, making thermal fluctuations more prominent. Consequently, PIC noise may enhance the overall perturbation growth rate.

\section{Results in the ultra-relativistic regime\label{sec:ur_results}}

\begin{figure*}
    \centering
    \includegraphics[width=1\linewidth]{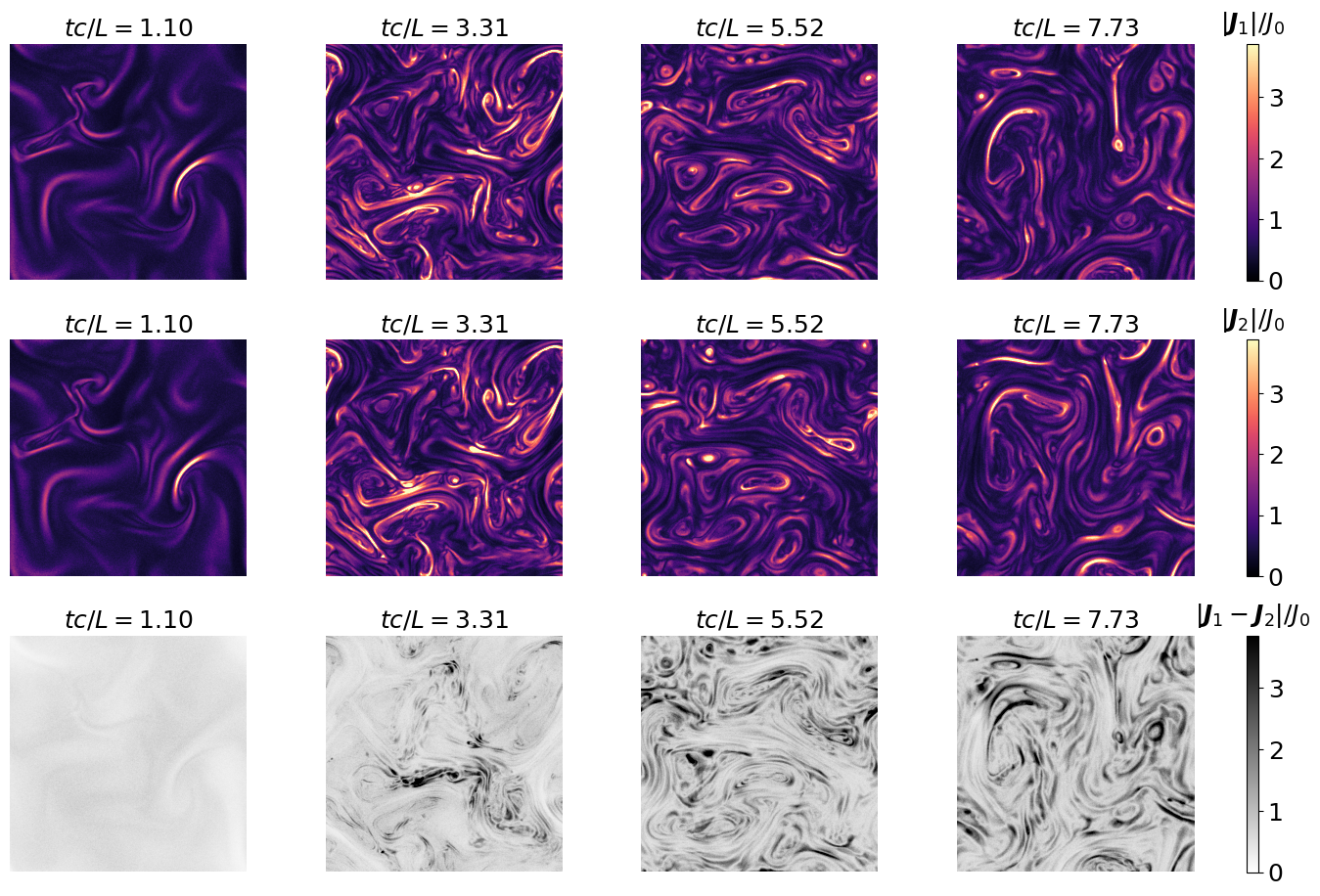}
    \caption{Evolution of the current density in the ultra-relativistic regime with $T_0/m_ec^2=10$, $\beta=1$, $\Delta\theta=\pi$, and $r_0=L/8192$. From top to bottom, the rows show the current density magnitude in the unperturbed (baseline) simulation, $|\boldsymbol{J}_1|$; the current density magnitude in the perturbed simulation, $|\boldsymbol{J}_2|$; and the magnitude of their difference, $|\Delta\boldsymbol{J}|=|\boldsymbol{J}_1-\boldsymbol{J}_2|$. The columns show different numbers of light-crossing times, $tc/L$, as indicated above the panels. All current densities are normalized by the characteristic current density $J_0$ defined in Sec.~\ref{sec:definitions}. The plots for $|\boldsymbol{J}_1|$ and $|\boldsymbol{J}_2|$ share a common linear color scale, which differs from that used for $|\Delta\boldsymbol{J}|$. In each scale, the lower limit is zero and the upper limit is the $99.5$th percentile of all values displayed in the corresponding rows; values above this limit are saturated.}
    \label{fig:rel_current_density_visual}
\end{figure*}

We now turn our attention to the evolution of simulations in the ultra-relativistic regime ($T_0/m_ec^2 \gg 1$) to explore whether there are any qualitative differences from the sub-relativistic cases. We will use light-crossing times instead of sound-crossing times from our previous sub-relativistic analysis, since the two become comparable: in particular, $c_s = (2/5)^{1/2} c$ for a collisionless ultra-relativistic pair plasma \citep{ugarov_etal_2026}. The fiducial temperature in the ultra-relativistic regime is $T_0/m_ec^2 = 10$, as opposed to $T_0/m_ec^2 = 0.05$ in Sec.~\ref{current_density_difference_section}. 
Fig.~\ref{fig:rel_current_density_visual} shows the current density magnitude in the perturbed simulation, $|\boldsymbol{J}_2|$, which diverges from that of the baseline simulation, $|\boldsymbol{J}_1|$. At the latest time displayed, $tc/L=7.73$, some large-scale similarities remain visible between $|\boldsymbol{J}_1|$ and $|\boldsymbol{J}_2|$, although their difference, $|\Delta\boldsymbol{J}|=|\boldsymbol{J}_1-\boldsymbol{J}_2|$, has spread throughout the domain. The figure is therefore intended to illustrate the development of finer visual differences as opposed to establishing the precise time at which the simulations become statistically decorrelated. The progression of the divergence broadly resembles that of the sub-relativistic regime in Fig.~\ref{current_density_visual}, with no apparent qualitative difference in the large-scale evolution.

\begin{figure*}
    \centering
    \includegraphics[width=1\linewidth]{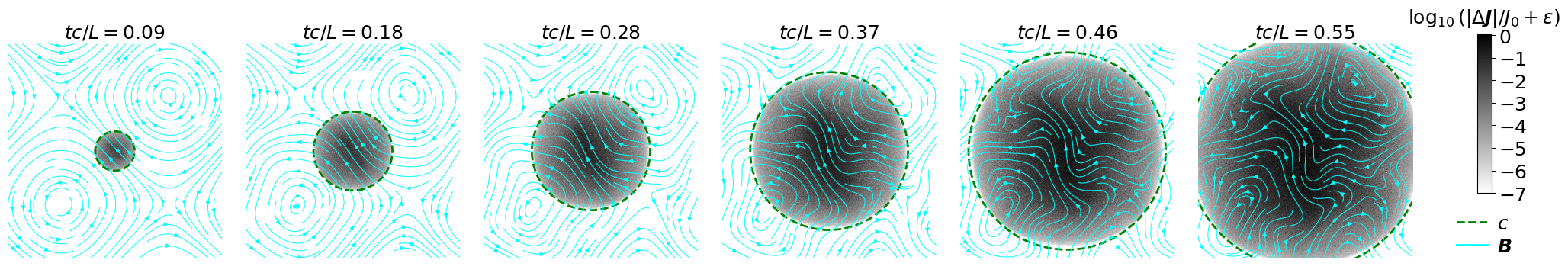}
    \caption{Early-time evolution of the total current density difference for the same ultra-relativistic case as in Fig.~\ref{fig:rel_current_density_visual}. The grayscale shows $\log_{10}(|\Delta\boldsymbol{J}|/J_0+\epsilon)$, where $|\Delta\boldsymbol{J}|=|\boldsymbol{J}_1-\boldsymbol{J}_2|$ and $\epsilon = 10^{-12}$ is a small parameter included solely to prevent taking the logarithm of zero. The dashed circle is centered on the initial perturbation, showing $r(t)=ct$ in green. The cyan streamlines represent the in-plane magnetic field $(B_x,B_y)$ from the baseline simulation. The times $tc/L$ are indicated above the panels.}
    \label{fig:rel_early_current_density_visual}
\end{figure*}

To further examine the spatial propagation of the perturbation, we observe the early-time evolution of $|\Delta\boldsymbol{J}|$ in Fig.~\ref{fig:rel_early_current_density_visual}. Compared with the sub-relativistic case in Fig.~\ref{early_current_density_visual}, the early perturbation appears more diffuse and has a less localized high-amplitude region (possibly because $c_s$ is comparable to $c$, blurring the distinction between high-amplitude and low-amplitude regions). In the sub-relativistic case, the localized structure was aligned with the in-plane magnetic-field lines, as discussed in Sec.~\ref{current_density_difference_section}. The more diffuse ultra-relativistic structure therefore suggests that the perturbation spreads across the field lines more efficiently in this regime, despite the similar plasma beta.

\begin{figure}
    \centering
    \includegraphics[width=1\linewidth]{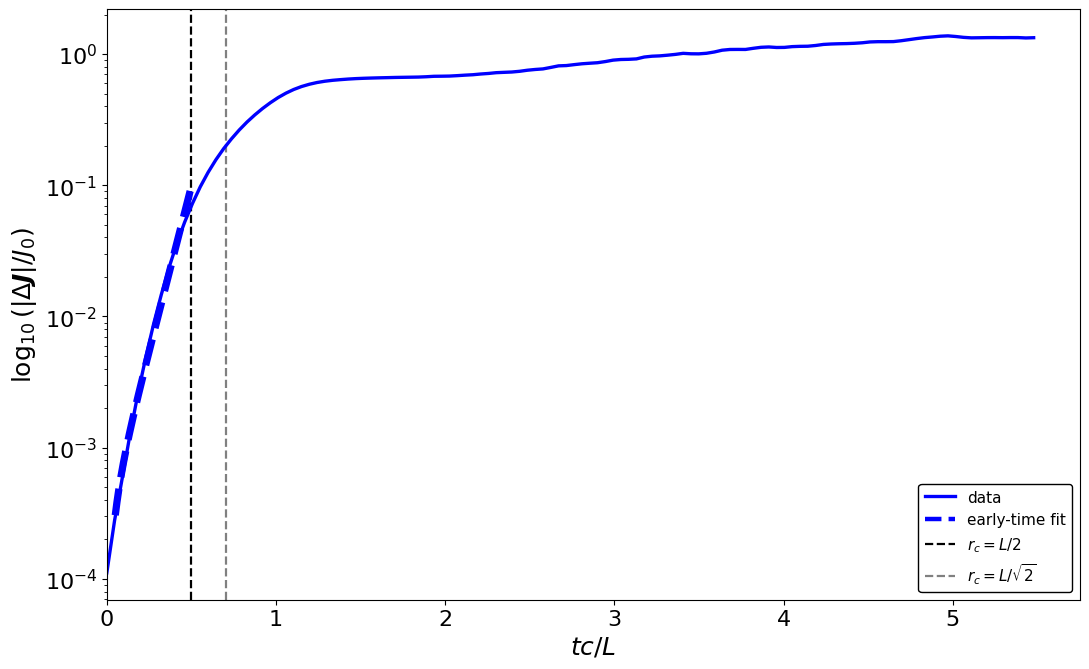}
    \caption{Evolution of the normalized, logarithmic rms total current density difference, $\log_{10}(|\Delta\boldsymbol{J}|_{\rm rms}/J_0)$, in the ultra-relativistic regime with $T_0/m_ec^2=10$, $\beta=1$, $\Delta\theta=\pi$, and $r_0 = L/8192$. The solid curve shows the simulation data, and the dashed curve shows a fit to the early-time model from Sec.~\ref{early_time_pert} over $0.05\leq tc/L\leq0.5$. The vertical black and gray dashed lines mark the times at which the information front, $r_c(t)=ct$, reaches $r_c=L/2$ and $r_c=L/\sqrt{2}$, respectively.}
    \label{fig:rel_current_early}
\end{figure}

Fig.~\ref{fig:rel_current_early} depicts the data for the ultra-relativistic $|\Delta\boldsymbol{J}|_{\rm rms}/J_0$ on a logarithmic vertical scale, and uses the early-time model from Sec.~\ref{early_time_pert} to generate the fit line. The displayed interval emphasizes the early perturbation growth and its transition into gradual saturation. At later times, this difference approaches saturation near $\sqrt{2}$, as expected from decorrelation. These times are omitted in order to better emphasize the early-time data. The best fit parameters are $J_{\rm char}/J_0=3.02\times10^{-4}$ and $\lambda=11.58c/L$. Since this value of $\lambda$ is defined with respect to $tc/L$, we can compare with the fiducial sub-relativistic value of $\lambda_c=2.20c/L$ reported in Sec.~\ref{temp}. Although the fitting intervals are not identical, the substantially larger fitted value in the ultra-relativistic case indicates more rapid early-time growth when both regimes are expressed in light-crossing units. This is consistent with the larger characteristic particle speeds in the ultra-relativistic plasma. 

The remaining parameter variation tests considered in the sub-relativistic analysis were also repeated in the ultra-relativistic regime. However, within the same parameter ranges, no qualitatively new results were produced beyond the trends already described in Sec.~\ref{sec:results}. We therefore do not reproduce the additional parameter scans here.

\section{Conclusions}

The purpose of this work was to investigate the butterfly effect within collisionless plasma turbulence. To do this, we performed and analyzed particle-in-cell simulations of electron-positron plasma to investigate the growth of initially localized perturbations. To characterize the early evolution of the difference between the simulations, we developed a model that combines the spatial expansion of the affected region with exponential amplification within that region. Perturbations with radii as small as $r_0=L/8192$, corresponding to a sub-cell-scale region containing only a few macroparticles per species, were found to grow and produce macroscopic differences throughout the simulation domain. In both the sub-relativistic and ultra-relativistic regimes, the paired simulations eventually reached statistical decorrelation in both current density and magnetic field. Demonstrating this behavior in a collisionless kinetic system extends the butterfly effect beyond conventional fluid turbulence, showing that particle-scale differences can also be amplified through plasma dynamics. In particular, our work implies that the butterfly effect is uninhibited by information loss through irreversible collective plasma processes.

In the fiducial sub-relativistic case, the highest-amplitude current density difference develops primarily within the sound-speed propagation front, with a much weaker difference extending outward near the information front (the region constrained by causality). The central part of the perturbation evolves into an anisotropic, snake-like structure that is aligned with the in-plane magnetic field lines. This suggests that the magnetic field influences the spatial development of the perturbation. Prior to the relevant propagation front reaching the periodic boundaries, our early-time model in Sec.~\ref{early_time_pert} reasonably describes the rms current density and magnetic field differences in the sub-relativistic regime, as well as the rms current density difference in the ultra-relativistic regime. 

Although the ultra-relativistic simulations can be modeled by the same early-time description as the sub-relativistic simulations, the actual early-time geometry appears much more diffuse than the localized, snake-like geometry of the sub-relativistic regime. A clear comparison between the growth rates of the two regimes shows that the ultra-relativistic regime has a substantially larger fitted growth rate under a common light-crossing time normalization. The initial perturbation scans demonstrate that the perturbation radius and rotation angle primarily determine the early evolution as opposed to the final outcome. Regardless of initial radius or rotation angle, all driven simulations eventually converge towards the same normalized saturation level, $|\Delta \boldsymbol{J}|_{\rm rms}/J_0\approx\sqrt{2}$, as expected of statistical decorrelation.

In the sub-relativistic regime, we performed a temperature scan where it was found that the fitted early-time growth rate subtly increases with temperature when compared using the same light-crossing time normalization. However, due to the limited temperature range and the deterioration of the fit at the lowest temperatures, we were unable to establish a quantitative temperature scaling. The current density and magnetic field growth rates also vary nonlinearly with plasma beta, although this dependence may be physical and/or numerical because the level of noise increases with beta. 

An important limitation of this work is that the contribution of numerical noise is artificially large compared to physical systems, which nearly always have many more particles. We explored this numerical contribution using two undriven control pairs. The control retaining the initial magnetic field fluctuations follows the driven case closely during much of its early evolution, whereas removing those initial fluctuations produces substantially slower growth, suggesting that the initialized magnetic structure plays an important role in the early amplification of the perturbation even without continued external driving. Relative to their own characteristic rms current scales, however, the controls still approach the saturation level associated with decorrelation. The lower absolute saturation level demonstrates the weaker overall amplification in the undriven simulations, while their eventual decorrelation suggests that numerical differences may still become significant over sufficiently long times.

There are several possible future extensions that could build on our results. To better separate the physical perturbation growth from the numerical contributions, such extensions could involve sampling a broader range of temperatures and plasma beta values to establish a quantitative scaling of the growth rate with plasma parameters, varying the number of particles per cell, and comparing to continuum Vlasov codes (which do not suffer from particle noise). Additional insights could come from more carefully analyzing the diffuse spatial structure found at early times in the ultra-relativistic regime. A more detailed understanding of this early stage could also be gained by tracking individual particle trajectories (in particular, the initially perturbed particles). Finally, the simulations could be extended to three dimensions and electron-ion plasmas. We conjecture that our analytical model can be extended to describe the early stages of expansion in such situations. Modifications will be required to account for 3D geometry (where the information front is spherical rather than circular), anisotropy with respect to the mean magnetic field, and additional characteristic expansion velocities associated with the different species.

These results provide strong numerical evidence for a butterfly effect within kinetic plasma turbulence, with initially localized, microscopic differences leading to macroscopically distinct plasma configurations. The eventual decorrelation was found across initial perturbation sizes, rotation angles, plasma beta values, and temperatures (including the ultra-relativistic regime), in both the plasma's current density and magnetic field, establishing a finite predictability horizon for systems differing only by a slight localized perturbation. While our present study pertains to kinetic turbulence, it is reasonable to conjecture that the conclusions may be more universal. For example, a similar butterfly effect could apply to other irreversible processes in collisionless plasmas, such as magnetic reconnection and shocks.

\begin{acknowledgments}
VZ would like to thank Clint Sprott for early discussions that helped inspire this project. VZ acknowledges support from the National Science Foundation under NSF grant PHY-2409316, and the Department of Energy under the grant DE-SC0026099. This work used Stampede3 at the Texas Advanced Computer Center (TACC) through allocation PHY160032 from the Advanced Cyberinfrastructure Coordination Ecosystem: Services \& Support (ACCESS) program, which is supported by U.S. National Science Foundation grants \#2138259, \#2138286, \#2138307, \#2137603, and \#2138296. 
\end{acknowledgments}
\nocite{*}
\bibliography{refs}

\end{document}